\newif\ifpeer
\documentclass[journal]{IEEEtran}

\usepackage[dvips]{graphicx}
\usepackage{amsmath,amssymb}
\usepackage{graphicx}
\usepackage{placeins}
\usepackage{wrapfig}
\usepackage{verbatim}
\usepackage{flushend}
\usepackage{subcaption}
\usepackage{tikz}
\usetikzlibrary{calc}
\usetikzlibrary{patterns}
\usepackage[ruled,vlined,boxed,algonl,linesnumbered]{algorithm2e}
\usepackage{ifoddpage}
\usepackage{etoolbox}
\usepackage{varioref}
\newtoggle{twocol}
\ifpeer
    \togglefalse{twocol}
\else
    \toggletrue{twocol}
\fi
\iftoggle{twocol}{\newcommand{\splitone}{}}{\newcommand{\splitone}{\nonumber \\ &}}
\newcommand{\splital}{\nonumber \\ &}
\iftoggle{twocol}{\newcommand{\splitwo}{\nonumber \\ &}}{\newcommand{\splitwo}{}}

\iftoggle{twocol}{\newcommand{\figwidth}{0.4}}{\newcommand{\figwidth}{}}

\iftoggle{twocol}{\newcommand{\onetable}[1]{
    \begin{table}
    \centering
    #1
    \end{table}
}}{\newcommand{\onetable}[1]{
    \linespread{1}
    \begin{table}
    \centering
    #1
    \end{table}
    \linespread{2}
}}

\iftoggle{twocol}{\newcommand{\pairtable}[2]{
    \onetable{#1}
    \onetable{#2} 
}}{\newcommand{\pairtable}[2]{
    \linespread{1}
    \begin{table}
    \parbox{.47\textwidth}{
    \centering
    #1
    }
    \hfill
    \parbox{.47\textwidth}{
    \centering
    #2
    }
    \end{table}
    \linespread{2}
}}

\iftoggle{twocol}{\newcommand{\onealg}[1]{%
\IncMargin{1em}%
\begin{algorithm}%
\DontPrintSemicolon%
#1
\end{algorithm}%
\DecMargin{1em}%
}}{%
\newcommand{\onealg}[1]{%
\linespread{1}%
\IncMargin{1em}%
\begin{algorithm}
\DontPrintSemicolon
#1
\end{algorithm}%
\DecMargin{1em}%
\linespread{2}}
}

\iftoggle{twocol}{\newcommand{\pairalg}[2]{
    \onealg{#1}
    \onealg{#2}}
    }{\newcommand{\pairalg}[2]{
    \IncMargin{1em}
    \linespread{1}
    \begin{figure*}
    \makebox[0.5\textwidth][l]{
        \small
        \begin{minipage}[t]{0.5\textwidth}
        \begin{algorithm}[H]
        \DontPrintSemicolon
        #1
        \end{algorithm}
        \end{minipage}
    }
    \makebox[0.5\textwidth][l]{
        \small
        \begin{minipage}[t]{0.5\textwidth}
        \begin{algorithm}[H]
        \DontPrintSemicolon
        #2
        \end{algorithm}
        \end{minipage}
    }
    \end{figure*}
    \DecMargin{1em}
    \linespread{2} 
}}

\iftoggle{twocol}{\newcommand{\pairalgr}[2]{
    \pairalg{#1}{#2}
}}{\newcommand{\pairalgr}[2]{
    \IncMargin{1em}
    \linespread{1}
    \begin{figure*}
    \makebox[0.4\textwidth][l]{
        \begin{minipage}{0.4\textwidth}
        \begin{algorithm}[H]
        \DontPrintSemicolon
        #1
        \end{algorithm}
        \end{minipage}
    }
    \makebox[0.6\textwidth][r]{
        \begin{minipage}{0.59\textwidth}
        \begin{algorithm}[H]
        \DontPrintSemicolon
        #2
        \end{algorithm}
        \end{minipage}
    }
    \end{figure*}
    \DecMargin{1em}
    \linespread{2} 
}}

\iftoggle{twocol}{}{}

\iftoggle{twocol}{\newcommand{\pairfig}[2]{
    \onefig{#1}
    \onefig{#2} 
}}{\newcommand{\pairfig}[2]{
   \begin{figure}
  \begin{minipage}[t]{0.48\textwidth}
   #1
  \end{minipage} 
  \hfill
  \begin{minipage}[t]{0.48\textwidth}
    #2
  \end{minipage}
\end{figure}

    \linespread{2}
}}

\definecolor{commentcolor}{rgb}{0,0,0.7}

\SetCommentSty{mycommfont}
\let\oldnl\nl
\newcommand{\nonl}{\renewcommand{\nl}{\let\nl\oldnl}}
\newcommand{\mA}{\mathcal{A}}
\newcommand{\mB}{\mathcal{B}}
\newcommand{\mC}{\mathcal{C}}

\newcommand{\mF}{\mathcal{F}}

\newcommand{\mI}{\mathcal{I}}

\newcommand{\mW}{\mathcal{W}}

\newcommand{\mY}{\mathcal{Y}}

\newcommand{\bF}{\mathbb{F}}

\newcommand{\bR}{\mathbb{R}}

\newcommand{\bfA}{\mathbf{A}}

\newcommand{\bfC}{\mathbf{C}}

\newcommand{\bfI}{\mathbf{I}}

\newcommand{\bfR}{\mathbf{R}}

\newcommand{\bfT}{\mathbf{T}}

\newcommand{\omC}{\overline{\mathcal{C}}}

\newcommand{\ve}{\varepsilon}

\newcommand{\tL}{\widetilde{L}}

\newcommand{\oL}{\overline L}
\newcommand{\omu}{\overline{\mu}}

\newcommand{\vp}{\varphi}

\newcommand{\osigma}{\overline{\sigma}}

\newcommand{\ox}{\overline{x}}
\newcommand{\tx}{\widetilde{x}}

\DeclareMathOperator{\sgn}{sign}

\newcommand{\set}[1]{\left\{{#1}\right\}}

\newcommand{\floor}[1]{\left\lfloor{#1}\right\rfloor}
\newcommand{\bnull}{\mathbf 0}

\newcommand{\call}[1]{\mbox{\texttt{#1}}}
\newcommand{\var}[1]{\mbox{\textsf{#1}}}

\newcommand{\mgets}{\overset{-}{\gets}}
\newcommand{\pgets}{\overset{+}{\gets}}
\newcommand{\opgets}{\overset{\oplus}{\gets}}

\newtheorem{remark}{Remark}

\begin{document}
\graphicspath{{./}}
\title{Efficient List Decoding of Convolutional Polar Codes}
\author{Ruslan Morozov\\ITMO University, Saint Petersburg, Russia\\rmorozov@itmo.ru}

\sloppy
\maketitle
\begin{abstract}
An efficient implementation of min-sum SC/list decoding of convolutional polar codes is proposed.
The complexity of the proposed implementation of SC decoding is more than two times smaller than the straightforward implementation.
Moreover, the proposed list decoding algorithm does not require to copy any LLRs during decoding.
\end{abstract}

\section{Introduction}
Convolutional polar codes (CvPCs, also known as b-MERA codes) \cite{ferris2017convolutional} are a family of linear block codes that employ channel polarization phenomenon \cite{arikan2009channel}.
Although there is no proof of capacity-achieving property and channel polarization, CvPCs are shown \cite{saber2018convolutional,morozov2018efficient,prinz2018successive} to perform better than Arikan polar codes under list decoding \cite{tal2015list}
with the same list size.
Note that the same list size for CvPCs require more arithmetic operations compared to Arikan polar codes.

Moreover, for sufficiently small error probabilities CvPCs even outperform Arikan polar codes by complexity, since for small error probability CvPCs need much smaller
list size than Arikan polar codes.
However, CvPCs outperform Arikan polar codes both by complexity and error-correcting performance only for very large list sizes, approximately a few hundred.
Such decoder complexity has little application in practice.

This paper is focused on detailed description of list decoding implementation for CvPCs. It requires one to substantially alter the way that the list decoder works with Tal-Vardy data structures.
We have two goals: 1) to propose an  efficient list decoding algorithm, and 2) for those, who are not closely familiar with CvPCs, to make them able to easily reproduce the list decoding algorithm for CvPCs, and then try to generalize and/or improve it further.

The paper is organized as follows.
In Section~\ref{s:encsc}, encoding and straightforward SC decoding implementation is given.
Then, in Section~\ref{s:sceff}, various optimizations are introduced, which reduce the complexity of SC decoding.
Finally, in Section~\ref{s:listeff}, an efficient list decoding implementation is given.

\section{Encoding and SC Decoding}
\label{s:encsc}
\subsection{Encoding}
An $(n=2^m,k)$ convolutional polar code (CvPC), $m\geq 2$, is defined as a set of vectors 
\begin{align}
c_{[n]}=u_{[n]}Q^{(n)}, u_{\mF}=\bnull^{n-k-1}, \mF\subset[n],
\end{align}
where $[n]$ denotes set $\{0,1,\ldots,n-1\}$, $|\mF|=n-k$ is the set of frozen symbols of $u$, and the remaining bits $u_{\mI}$, where $\mI=[n]\setminus\mF$, carry payload data $a_{[k]}$.
Here and throughout the paper, $a_{\mB}$ denotes vector of elements of $a$ with indices from set $\mB$, so e.g. $a_{[n]}=a_0^{n-1}$.

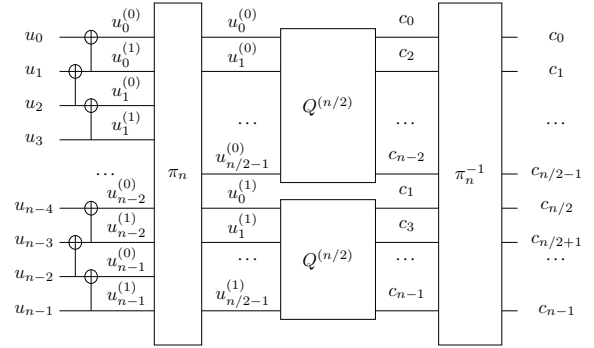
\begin{figure}
\centering
\iftoggle{twocol}{\scalebox{0.7}{\begin{tikzpicture}[x=0.6cm,y=0.65cm]
\def\r{0.2}
\def\g{0.5}


\foreach \x in {1,...,4,6,7,...,9}{
  \draw (1,\x)--(4,\x);
}
\foreach \x in {1,3,4,5,8,9}{
\draw (5.5,\x)--(8,\x);
}
\foreach \x in {1,3,4,5,8,9}{
\draw (11,\x)--(13,\x);
}
\foreach \x in {1,3,4,5,8,9}{
\draw (15,\x)--(15.5,\x);
}


\foreach \x in {2,7}{
\draw (1.5,\x)--(1.5,\x+1+\r);
\draw (1.5,\x+1) circle (\r);
}

\foreach \x in {1,3,8,6}{
\draw (2,\x)--(2,\x+1+\r);
\draw (2,\x+1) circle (\r);
}


\draw (4,0) rectangle (5.5,10);
\draw (8,0.75) rectangle (11,4.25);
\draw (8,4.75) rectangle (11,9.25);
\draw (13,0) rectangle (15,10);


\node at (2.5,5) {\ldots};
\node at (7,6.5) {\ldots};
\node at (7,2.5) {\ldots};
\node at (12,6.5) {\ldots};
\node at (12,2.5) {\ldots};
\node at (16.75,6.5) {\ldots};
\node at (16.75,2.5) {\ldots};

\node at (4.75,5) {$\pi_n$};
\node at (14,5) {$\pi^{-1}_{n}$};

\foreach \x in {0,...,3}{
  \node at (0.2,9-\x) {$u_\x$};
}
\foreach \x in {0,1}{
\node at (3.125,9+\g-2*\x) {$u^{(0)}_\x$};
\node at (3.125,8+\g-2*\x) {$u^{(1)}_\x$};
}
\foreach \x in {1,...,4}{
  \node at (0.2,\x) {$u_{n-\x}$};
}
\foreach \x in {1,2}{
\node at (3.125,-1+\g+2*\x) {$u^{(1)}_{n-\x}$};
\node at (3.125,\g+2*\x) {$u^{(0)}_{n-\x}$};
}

\def\xx{6.8}
\node at (\xx,1+\g) {$u^{(1)}_{n/2-1}$};
\node at (\xx,3+\g) {$u^{(1)}_{1}$};
\node at (\xx,4+\g) {$u^{(1)}_{0}$};
\node at (\xx,5+\g) {$u^{(0)}_{n/2-1}$};
\node at (\xx,8+\g) {$u^{(0)}_1$};
\node at (\xx,9+\g) {$u^{(0)}_0$};

\node at (9.5,7) {$Q^{(n/2)}$};
\node at (9.5,2.5) {$Q^{(n/2)}$};

\node at (12,1+\g) {$c_{n-1}$};
\node at (12,3+\g) {$c_{3}$};
\node at (12,4+\g) {$c_{1}$};
\node at (12,5+\g) {$c_{n-2}$};
\node at (12,8+\g) {$c_2$};
\node at (12,9+\g) {$c_0$};

\node at (16.75,1) {$c_{n-1}$};
\node at (16.75,3) {$c_{n/2+1}$};
\node at (16.75,4) {$c_{n/2}$};
\node at (16.75,5) {$c_{n/2-1}$};
\node at (16.75,8) {$c_1$};
\node at (16.75,9) {$c_0$};
\end{tikzpicture}}}{\begin{tikzpicture}[x=0.6cm,y=0.65cm]
\def\r{0.2}
\def\g{0.5}


\foreach \x in {1,...,4,6,7,...,9}{
  \draw (1,\x)--(4,\x);
}
\foreach \x in {1,3,4,5,8,9}{
\draw (5.5,\x)--(8,\x);
}
\foreach \x in {1,3,4,5,8,9}{
\draw (11,\x)--(13,\x);
}
\foreach \x in {1,3,4,5,8,9}{
\draw (15,\x)--(15.5,\x);
}


\foreach \x in {2,7}{
\draw (1.5,\x)--(1.5,\x+1+\r);
\draw (1.5,\x+1) circle (\r);
}

\foreach \x in {1,3,8,6}{
\draw (2,\x)--(2,\x+1+\r);
\draw (2,\x+1) circle (\r);
}


\draw (4,0) rectangle (5.5,10);
\draw (8,0.75) rectangle (11,4.25);
\draw (8,4.75) rectangle (11,9.25);
\draw (13,0) rectangle (15,10);


\node at (2.5,5) {\ldots};
\node at (7,6.5) {\ldots};
\node at (7,2.5) {\ldots};
\node at (12,6.5) {\ldots};
\node at (12,2.5) {\ldots};
\node at (16.75,6.5) {\ldots};
\node at (16.75,2.5) {\ldots};

\node at (4.75,5) {$\pi_n$};
\node at (14,5) {$\pi^{-1}_{n}$};

\foreach \x in {0,...,3}{
  \node at (0.2,9-\x) {$u_\x$};
}
\foreach \x in {0,1}{
\node at (3.125,9+\g-2*\x) {$u^{(0)}_\x$};
\node at (3.125,8+\g-2*\x) {$u^{(1)}_\x$};
}
\foreach \x in {1,...,4}{
  \node at (0.2,\x) {$u_{n-\x}$};
}
\foreach \x in {1,2}{
\node at (3.125,-1+\g+2*\x) {$u^{(1)}_{n-\x}$};
\node at (3.125,\g+2*\x) {$u^{(0)}_{n-\x}$};
}

\def\xx{6.8}
\node at (\xx,1+\g) {$u^{(1)}_{n/2-1}$};
\node at (\xx,3+\g) {$u^{(1)}_{1}$};
\node at (\xx,4+\g) {$u^{(1)}_{0}$};
\node at (\xx,5+\g) {$u^{(0)}_{n/2-1}$};
\node at (\xx,8+\g) {$u^{(0)}_1$};
\node at (\xx,9+\g) {$u^{(0)}_0$};

\node at (9.5,7) {$Q^{(n/2)}$};
\node at (9.5,2.5) {$Q^{(n/2)}$};

\node at (12,1+\g) {$c_{n-1}$};
\node at (12,3+\g) {$c_{3}$};
\node at (12,4+\g) {$c_{1}$};
\node at (12,5+\g) {$c_{n-2}$};
\node at (12,8+\g) {$c_2$};
\node at (12,9+\g) {$c_0$};

\node at (16.75,1) {$c_{n-1}$};
\node at (16.75,3) {$c_{n/2+1}$};
\node at (16.75,4) {$c_{n/2}$};
\node at (16.75,5) {$c_{n/2-1}$};
\node at (16.75,8) {$c_1$};
\node at (16.75,9) {$c_0$};
\end{tikzpicture}}
\caption{Recursive structure of CvPT $Q^{(n)}$ of size $n$}
\label{fig:cvpt}
\end{figure}

The convolutional polarizing transformation (CvPT) $Q^{(n)}$ is an $n \times n$ matrix, which has the recursive structure, depicted in Fig.~\ref{fig:cvpt}, and
is given by the recursion
\begin{align}
Q^{(n)}=(X^{(n)}Q^{(n/2)},Z^{(n)}Q^{(n/2)})\pi^{-1}_{n},
\label{eq:cvpt}
\end{align}
where $\pi_n$ is the matrix of permutation ``even first odd last'' $(0,2,\ldots,n-2,1,3,\ldots,n-1)$, $Q^{(1)}=(1)$, $X^{(l)}$ and $Z^{(l)}$ are $l\times l/2$ matrices, defined for even $l$ as
\begin{align}
\label{eq:xzdef}
X^{(l)}_{i,j} =\boldsymbol1[2j\leq i \leq 2j+2],
Z^{(l)}_{i,j} =\boldsymbol1[2j< i \leq 2j+2].
\end{align}
where $\mathbf{1}[\text{statement}]=1$, if the statement is true, and $0$  otherwise. For example, 
$X^{(4)}=\begin{pmatrix}1110\\0011\end{pmatrix}^T$,
$Z^{(4)}=\begin{pmatrix}0110\\0001\end{pmatrix}^T$.
Expansion~\eqref{eq:cvpt} corresponds to one \textit{layer} of the CvPT.
In Fig.~\ref{fig:cvpt}, the $m$-th layer of the CvPT is a mapping of vector $u_{[n]}$ onto vectors $u^{(0)}_{[n/2]}=u_{[n]}X^{(n)}$ and $u^{(1)}_{[n/2]}=u_{[n]}Z^{(n)}$.
Thus, a codeword $c_{[n]}$ can be expressed as  $c_{[n]}=u_{[n]}Q^{(n)}=(u^{(0)}_{[n/2]}Q^{(n/2)},u^{(1)}_{[n/2]}Q^{(n/2)})\pi_n^{-1}$,
where
\begin{align}
u^{(0)}_i=u_{2i}\oplus u_{2i+1}\oplus u_{2i+2}, \;&u^{(1)}_i=u_{2i+1}\oplus u_{2i+2}, i\leq \frac{n}{2}\!-\!2, \nonumber\\
u^{(0)}_{n/2-1}=u_{n-2}\oplus u_{n-1}, \;&u^{(1)}_{n/2-1}=u_{n-1}.
\label{eq:u01}
\end{align}
Applying $^{(0)}$ and $^{(1)}$ operations to $u^{(0)}$ and $u^{(1)}$ one can obtain $u^{(00)}_{[n/4]}$, $u^{(10)}_{[n/4]}$, $u^{(01)}_{[n/4]}$,
$u^{(11)}_{[n/4]}$, etc.



\subsection{Successive Cancellation Decoding}
\label{ss:scgen}
In this section we introduce the min-sum version of  successive cancellation (SC) decoding algorithm for CvPC \cite{morozov2018efficient}.
We also introduce the notation that will be used throughout the paper.

Consider transmission of codeword $c_{[n]}= u_{[n]}Q^{(n)}$ through binary-input memoryless channel $\mW:\bF\to\mY$, where $\bF$
is the binary field, $\mY$ is the output alphabet of $\mW$.
Let $y_{[n]}\in\mY^n$ be the channel output.
The demodulated probabilities $W(c_i|y_i)=\mW(y_i|c_i)/\left(\mW(y_i|0)+\mW(y_i|1)\right)$ for $c_i \in \bF$ are provided to the decoding algorithm.

Define \textit{a} $t$-\textit{cluster} $A$ as an array of $2^t$ real values, which are indexed by $t$ bits $A[x_0,x_1,...,x_{t-1}]=A[x_{[t]}]$, $x_i\in\bF$.

Assuming some fixed prior hard decisions $\hat u_{[\vp]}$ on first
$\vp$ symbols $u_{[\vp]}$, define the log-likelihood of vector $x_{[t]}$
for any $0\leq t<n-\vp$ as a $t$-cluster 
\begin{align}
L_\vp[x_{[t]}]=\ln\max_{x_{t}^{n-\vp-1}\in\bF^{n-\vp-t}}W^n\left((\hat u_{[\vp]},x_{[n-\vp]})Q^{(n)}|y_{[n]}\right),
\label{eq:ldef}
\end{align}
where $W^n(c_{[n]}|y_{[n]})=\prod_{i=0}^{n-1}W(c_i|y_i)$.
Define also the log-likelihood ratio (LLR) of symbol $u_\vp$ as 
\begin{align}
\tL_\vp=L_\vp[0]-L_\vp[1],
\label{eq:tldef}
\end{align}
where $L_\vp$ is a $1$-cluster.
At the $\vp$-th phase the min-sum SC decoding algorithm computes value of $\tL_\vp$. Then, the hard decision on  $u_{\vp}$ is made by
\begin{align}
\hat u_\vp=\mathbf{1}[\vp \in \mI \wedge \tL>0].
\label{eq:hd}
\end{align}
It is shown in \cite{morozov2018efficient} that the value of 3-clusters $L_\vp[x_0,x_1,x_2]$ and 2-clusters $L_\vp[x_0,x_1]$ are given by recursion
\begin{align}
&L_0[x_{[2]}]=\max_{x_2^3\in\bF^2}\{L^{(0)}_0[x_{[4]}X^{(4)}]+L_0^{(1)}[x_{[4]}Z^{(4)}]\}
\splital
L_0[x_{[3]}]=\max_{x_3\in\bF}\{L^{(0)}_0[x_{[4]}X^{(4)}]+L_0^{(1)}[x_{[4]}Z^{(4)}]\}
\splital
L_{2\psi+1}[x_1^3]\!=\!\max_{x_4^5\in\bF^2,x_0=\hat u_{2i}}\{L^{(0)}_{\psi}\![x_{[6]}X^{(6)}]
\!+\!L_\psi^{(1)}\![x_{[6]}Z^{(6)}]\}
\splital
L_{2\psi+2}[x_2^4]\!=\!\max_{x_5\in\bF,x_0^1=\hat u_{2i}^{2i+1}}\{L^{(0)}_{\psi}\![x_{[6]}X^{(6)}]
+L^{(1)}_{\psi}\![x_{[6]}Z^{(6)}]\}
\splital
L_{n-3}[x_1^3]=_{(x_0=\hat u_{n-4})}L^{(0)}_{\frac{n}{2}-2}[x_{[4]}X^{(4)}]\!+\!L^{(1)}_{\frac{n}{2}-2}[x_{[4]}Z^{(4)}]
\splital
L_{n-3}[x_2^3]=_{(x_0^1=\hat u_{n-4}^{n-3})}L^{(0)}_{\frac{n}{2}-2}[x_{[4]}X^{(4)}]\!+\!L^{(1)}_{\frac{n}{2}-2}[x_{[4]}Z^{(4)}]
\label{eq:recl}
\end{align}
where subscript under $=$ sign is given to define the relationship between formal variables $x_i$ and hard decisions $\hat u_j$, 
\begin{align}
&L^{(0)}_\psi[x_{[t]}]\!=\ln\!\!\!\!\max_{x_{t}^{n/2-\psi-1}\in\bF^{\frac{n}{2}-\psi-t}}\!\!\!W^{\frac{n}{2}}((\hat
u^{(0)}_{[\psi]}, x_{[\frac{n}{2}-\psi]})Q^{(n/2)}|y^{(0)}) \nonumber\\
&L^{(1)}_\psi[x_{[t]}]\!=\ln\!\!\!\!\max_{x_{t}^{n/2-\psi-1}\in\bF^{\frac{n}{2}-\psi-t}}\!\!\!W^{\frac{n}{2}}((\hat
u^{(1)}_{[\psi]}, x_{[\frac{n}{2}-\psi]})Q^{(n/2)}|y^{(1)}), \label{eq:l01def}
\end{align}
$y^{(0)}=(y_0,y_2,...,y_{n-2})$, $y^{(1)}=(y_1,y_3,...,y_{n-1})$ are vectors of length $n/2$, and $\hat u^{(0)}$ and $\hat u^{(1)}$ are hard decisions on $u^{(0)}$ and $u^{(1)}$ on layer $m-1$, propagated from layer $m$ by~\eqref{eq:u01}.
Employing $^{(0)}$ and $^{(1)}$ operations to $y^{(0)}$,
$y^{(1)}$, $\hat u^{(0)}$ and $\hat u^{(1)}$, by~\eqref{eq:recl} one can express $L_\vp^{(0)}$, $L_\vp^{(1)}$ through $L_\psi^{(00)}$, $L_\psi^{(01)}$, $L_\psi^{(10)}$, $L_\psi^{(11)}$, etc.
The base of the recursion corresponds to the first layer of the CvPT.
The log-likelihoods on this layer for symbols $s_{[m-1]}\in\bF^{m-1}$ are
\begin{align}
L_0^{(s_{[m-1]})}[a,b]\!=\!\ln\left[W(a\!\oplus \!b|y_{J(s_{[m-1]},0)})W(b|y_{J(s_{[m-1]},1)})\right],
\label{eq:lrecbase}
\end{align}
where
\begin{align}
J(s_{[m]})=\sum_{j\in[m]}s_j2^j.
\label{eq:jdef}
\end{align}

One can represent all operations involved in~\eqref{eq:recl} as a special case of the following cluster operator.
Consider input $t'$-clusters $A,B$.
Then, the cluster operator $\sigma_{i,t,j}:\bR^{t'}\times\bR^{t'}\times \bF^i\to\bR^t$, $t=2t'-i-j$ for input bits $x_{[i]}\in\bF^i$ is defined as
\begin{align}
&\sigma_{i,t,j}(A,B,x_{[i]})=C: \text{  }
\splitwo
C[x_{i}^{i+t-1}]=\max_{x_{i+t}^{2t'-1}}\set{A[x_{[2t']}X^{(2t')}]+B[x_{[2t']}Z^{(2t')}]},
\label{eq:sigmadef}
\end{align}
where $2t'\times t'$ matrices $X^{(2t')}$ and $Z^{(2t')}$ are defined by~\eqref{eq:xzdef}.
\begin{figure}
\iftoggle{twocol}
    {
    \newcommand{\opwidth}{0.23\textwidth}\newcommand{\opx}{0.5cm}
    }{
    \newcommand{\opwidth}{0.32\textwidth}\newcommand{\opx}{0.55cm}
}
\newcommand{\opy}{\opx}
\newcommand{\rightx}{5.5}

\begin{subfigure}{\opwidth}
  \centering
\begin{tikzpicture}[x=\opx,y=\opy]
\def\r{0.1cm}
\draw (1,4)--(\rightx,4);
\draw (1,3)--(4,3)--(5,2)--(\rightx,2);
\draw (1,2)--(4,2)--(5,3)--(\rightx,3);
\draw (1,1)--(\rightx,1);

\draw (2,3) circle(\r);
\draw (2,2)--(2,3.2);
\draw (3,2) circle(\r);
\draw (3,3)--(3,4.2);
\draw (3,4) circle(\r);
\draw (3,1)--(3,2.2);

\draw (-0.3,2.8) rectangle (1,4.2);
\node at (0.4,3.5) {$L_0$};

\node at (1.3,4.3) {$a$};
\node at (1.3,3.3) {$b$};
\node at (1.3,2.3) {$v$};
\node at (1.3,1.3) {$w$};
\node at (0.3,2.2) {$\max$};
\node at (0.3,1.2) {$\max$};

\draw (\rightx,2.8) rectangle (\rightx+1.5,4.2);
\node at (\rightx+.75,3.5) {$L_0^{(0)}$};
\draw (\rightx,0.8) rectangle (\rightx+1.5,2.2);
\node at (\rightx+.75,1.5) {$L_0^{(1)}$};
\end{tikzpicture}
\caption{$\sigma_{0,2,2}$}
\label{fig:s022}
\end{subfigure}
\begin{subfigure}{\opwidth}
  \centering
\begin{tikzpicture}[x=\opx,y=\opy]
\def\r{0.1cm}
\draw (1,4)--(\rightx,4);
\draw (1,3)--(4,3)--(5,2)--(\rightx,2);
\draw (1,2)--(4,2)--(5,3)--(\rightx,3);
\draw (1,1)--(\rightx,1);

\draw (2,3) circle(\r);
\draw (2,2)--(2,3.2);
\draw (3,2) circle(\r);
\draw (3,3)--(3,4.2);
\draw (3,4) circle(\r);
\draw (3,1)--(3,2.2);

\draw (-0.3,1.8) rectangle (1,4.2);
\node at (0.4,3) {$L_0$};
\node at (1.3,2.3) {$c$};
\node at (1.3,3.3) {$b$};
\node at (1.3,4.3) {$a$};
\node at (1.3,1.3) {$v$};
\node at (0.3,1.2) {$\max$};
\draw (\rightx,2.8) rectangle (\rightx+1.5,4.2);
\node at (\rightx+.75,3.5) {$L_{0}^{(0)}$};
\draw (\rightx,0.8) rectangle (\rightx+1.5,2.2);
\node at (\rightx+.75,1.5) {$L_{0}^{(1)}$};
\end{tikzpicture}
\caption{$\sigma_{0,3,1}$}
\label{fig:s031}
\end{subfigure}
\iftoggle{twocol}{\\}{}
\begin{subfigure}{\opwidth}
\centering
\begin{tikzpicture}[x=\opx,y=\opy]
\def\r{0.1cm}
\draw (1,4)--(\rightx,4);
\draw (1,3)--(4,3)--(5,2)--(\rightx,2);
\draw (1,2)--(4,2)--(5,3)--(\rightx,3);
\draw (1,1)--(\rightx,1);

\draw (2,3) circle(\r);
\draw (2,2)--(2,3.2);
\draw (3,2) circle(\r);
\draw (3,3)--(3,4.2);
\draw (3,4) circle(\r);
\draw (3,1)--(3,2.2);

\draw (-0.8,0.8) rectangle (1,3.2);
\node at (0.15,2) {$L_{n-3}$};
\draw (\rightx,2.8) rectangle (\rightx+1.8,4.2);
\node at (\rightx+.9,3.5) {$L_{\frac{n}{2}-2}^{(0)}$};
\draw (\rightx,0.8) rectangle (\rightx+1.8,2.2);
\node at (\rightx+.9,1.5) {$L_{\frac{n}{2}-2}^{(1)}$};
\node at (1.3,1.3) {$c$};
\node at (1.3,2.3) {$b$};
\node at (1.3,3.3) {$a$};
\node at (0.2,4) {$\hat u_{n-4}$};
\end{tikzpicture}
  \caption{$\sigma_{1,3,0}$}
  \label{fig:s130}
\end{subfigure}
\begin{subfigure}{\opwidth}
  \centering
\begin{tikzpicture}[x=\opx,y=\opy]
\def\r{0.1cm}
\draw (1,4)--(\rightx,4);
\draw (1,3)--(4,3)--(5,2)--(\rightx,2);
\draw (1,2)--(4,2)--(5,3)--(\rightx,3);
\draw (1,1)--(\rightx,1);

\draw (2,3) circle(\r);
\draw (2,2)--(2,3.2);
\draw (3,2) circle(\r);
\draw (3,3)--(3,4.2);
\draw (3,4) circle(\r);
\draw (3,1)--(3,2.2);

\draw (-0.8,0.8) rectangle (1,2.2);
\node at (0.15,1.5) {$L_{n-2}$};
\draw (\rightx,2.8) rectangle (\rightx+1.8,4.2);
\node at (\rightx+.9,3.5) {$L_{\frac{n}{2}-2}^{(0)}$};
\draw (\rightx,0.8) rectangle (\rightx+1.8,2.2);
\node at (\rightx+.9,1.5) {$L_{\frac{n}{2}-2}^{(1)}$};
\node at (1.3,1.3) {$b$};
\node at (1.3,2.3) {$a$};
\node at (0.2,4) {$\hat u_{n-4}$};
\node at (0.2,3) {$\hat u_{n-3}$};
\end{tikzpicture}
\caption{$\sigma_{2,2,0}$}
\label{fig:s220}
\end{subfigure}
\iftoggle{twocol}{\\}{}
\begin{subfigure}{\opwidth}
\centering
\begin{tikzpicture}[x=\opx,y=\opy]
\def\r{0.1cm}
\draw (1,6)--(\rightx,6);
\draw (1,5)--(4,5)--(5,3)--(\rightx,3);
\draw (1,4)--(4,4)--(5,5)--(\rightx,5);
\draw (1,3)--(4,3)--(5,2)--(\rightx,2);
\draw (1,2)--(4,2)--(5,4)--(\rightx,4);
\draw (1,1)--(\rightx,1);

\draw (2,5) circle(\r);
\draw (2,3) circle(\r);
\draw (3,6) circle(\r);
\draw (3,4) circle(\r);
\draw (3,2) circle(\r);

\draw (2,2)--(2,3.2);
\draw (2,4)--(2,5.2);
\draw (3,1)--(3,2.2);
\draw (3,3)--(3,4.2);
\draw (3,5)--(3,6.2);

\draw (\rightx,3.8) rectangle (\rightx+1.5,6.2);
\node at (\rightx+.75,5) {$L_\psi^{(0)}$};
\draw (\rightx,0.8) rectangle (\rightx+1.5,3.2);
\node at (\rightx+.75,2) {$L_\psi^{(1)}$};
\node at (1.3,1.3) {$w$};
\node at (1.3,2.3) {$v$};
\node at (0.3,2.2) {$\max$};
\node at (0.3,1.2) {$\max$};
\node at (0.4,6) {$\hat u_{2\psi}$};

\node at (1.3,3.3) {$c$};
\node at (1.3,4.3) {$b$};
\node at (1.3,5.3) {$a$};

\draw (-1,2.8) rectangle (1,5.2);
\node at (0,4) {$L_{2\psi+1}$};
\end{tikzpicture}
\caption{$\sigma_{1,3,2}$}
\label{fig:s132}
\end{subfigure}
\begin{subfigure}{\opwidth}
 \centering
\begin{tikzpicture}[x=\opx,y=\opy]
\def\r{0.1cm}
\draw (1,6)--(\rightx,6);
\draw (1,5)--(4,5)--(5,3)--(\rightx,3);
\draw (1,4)--(4,4)--(5,5)--(\rightx,5);
\draw (1,3)--(4,3)--(5,2)--(\rightx,2);
\draw (1,2)--(4,2)--(5,4)--(\rightx,4);
\draw (1,1)--(5.5,1);

\draw (2,5) circle(\r);
\draw (2,3) circle(\r);
\draw (3,6) circle(\r);
\draw (3,4) circle(\r);
\draw (3,2) circle(\r);

\draw (2,2)--(2,3.2);
\draw (2,4)--(2,5.2);
\draw (3,1)--(3,2.2);
\draw (3,3)--(3,4.2);
\draw (3,5)--(3,6.2);

\draw (\rightx,3.8) rectangle (\rightx+1.5,6.2);
\node at (\rightx+.75,5) {$L_\psi^{(0)}$};
\draw (\rightx,0.8) rectangle (\rightx+1.5,3.2);
\node at (\rightx+.75,2) {$L_\psi^{(1)}$};
\node at (1.3,1.3) {$v$};
\node at (0.3,1.2) {$\max$};

\node at (0.1,5) {$\hat u_{2\psi+1}$};
\node at (0.4,6) {$\hat u_{2\psi}$};

\draw (-1,1.8) rectangle (1,4.2);
\node at (0,3) {$L_{2\psi+2}$};
\node at (1.3,2.3) {$c$};
\node at (1.3,3.3) {$b$};
\node at (1.3,4.3) {$a$};
\end{tikzpicture}
  \caption{$\sigma_{2,3,1}$}
  \label{fig:s231}
\end{subfigure}
\caption{Cluster operators used in min-sum SC decoding of CvPC}
\label{fig:patternssf}
\end{figure}
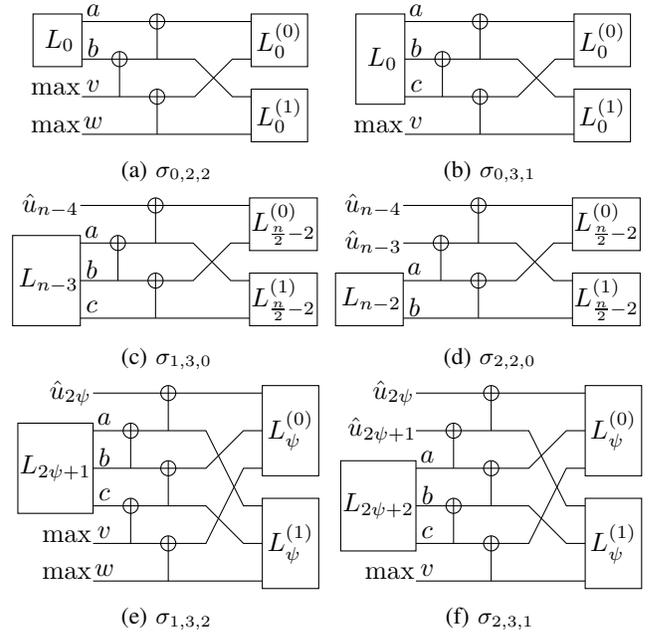
In Fig.~\ref{fig:patternssf}, all cluster operations, given by~\eqref{eq:recl}, are depicted.
The corresponding operator $\sigma_{i,t,j}$ is provided under each figure.
Input and output $2$-clusters and $3$-clusters are drawn by boxes with two or three inputs, corresponding to bits of index in these arrays, represented as binary vectors (the
first and least significant bit is on top).

For input $t'$-cluster $A$, and for input vector of $i$ bits $x_{[i]}$ define the marginalizing operator $\mu_{i,t,j}:\bR^{t'}\times \bF^i\to\bR^t$, $t=t'-i-j$, as
\begin{align}
&\mu_{i,t,j}(A,x_{[i]})=C: C[x_{i}^{i+t-1}]=\max_{x_{i+t}^{t'-1}\in\bF^j}\set{A[x_{[t']}]}.
\label{eq:mudef}
\end{align}
It can be seen that $L_\vp(u_\vp)=\mu_{0,1,t-1}(L_\vp(u_{\vp}^{\vp+t-1}))$.

For the case of $t'=1$, for $(i+j+1)$-cluster $A$ denote 
\begin{align}
\Delta\mu_{i,1,j}(A,x_{[i]})=\mu_{i,1,j}(A,x_{[i]})[0]-\mu_{i,1,j}(A,x_{[i]})[1].
\label{eq:dmudef}
\end{align}
Then, the values of $3$-clusters $L_\vp[u_\vp^{\vp+2}]$, obtained by~\eqref{eq:recl}, can be converted to LLR~\eqref{eq:tldef} as $\tL_\vp=\Delta\mu_{0,1,2}(L_\vp)$, $\vp\leq
n-3$.
For the last two phases one can employ
\begin{align}
\tL_{n-2}=\Delta\mu_{1,1,1}(L_{n-3},\hat u_{n-3}), \tL_{n-1}=\Delta\mu_{2,1,0}(L_{n-3},\hat u_{n-3}^{n-2}).
\label{eq:tllast}
\end{align}

\begin{remark}
\label{rm:clusteradd}
Assume $\oL^{(s,0)}[x_{[t]}]=L^{(s,0)}_{\psi}[x_{[t]}]+C_0$, $\oL^{(s,1)}_{\psi}[x_{[t]}]=L^{(s,1)}[x_{[t]}]+C_1$, where $L_\psi^{(s,b)}$, $b\in\bF$ are clusters on layer
$\lambda-1$, and $C_0^1$ does not depend on $x_{[t]}$.
Consider cluster $L_\vp^{(s)}$ on layer $\lambda$, which is computed from the clusters on the previous layer: $L_\vp^{(s)}=\sigma_{i,t',j}(L^{(s,0)}_\psi,L^{(s,1)}_\psi)$.
Replace the clusters with their shifted versions and denote $\oL_\vp^{(s)}=\sigma_{i,t',j}(\oL^{(s,0)}_\psi,\oL^{(s,1)}_\psi)$.
Then, $\oL^{(s)}_\vp[x_{[t']}]=L^{(s)}_\vp[x_{[t']}]+C_0+C_1$, and $C_0+C_1$ clearly does not depend on $x_{[t']}$.
Finally, on the layer $m$, we do not care about additive terms, since in the end we return the difference \eqref{eq:tldef}.
Thus, we need only to compute each cluster up to a constant.
\end{remark}


\subsection{Straightforward Implementation of SC Decoding}
\label{ss:scsf}

In Alg.~\ref{alg:calctsc}--\ref{alg:decodesc} the SC decoder implementation \cite{morozov2018efficient} is given.
For these algorithms, the parameters $m, n=2^m$, as well as arrays which are printed in bold ($\bfT$ and $\bfC$) are considered global.

Array $\bfT$ is a 3D array, indexed as $\bfT[\lambda][i][a,b,c]$, where $1\leq \lambda \leq m$, $i\in[2^{m-\lambda}]$, $(a,b,c)\in\bF^3$.
Thus, each array $\bfT[\lambda][i]$ is of size $8$ and represents $3$-cluster $L_\vp^{(s_{[m-\lambda]})}[a,b,c]$, where $i=J(s_{[m-\lambda]})$, $J$ is defined in \eqref{eq:jdef}
and phase $\vp$ is changing from $0$ to $2^{\lambda}-2$.
For $\lambda=1$, one can allocate $2$-clusters instead of $3$-clusters.
Array of pointers $\bfT[0]$ can be left uninitialized.

Array $\bfC$ is a 2D bit array of hard decisions, indexed as $\bfC[\lambda][2i+b]$, where $0\leq \lambda \leq m$, $i\in[2^{m-\lambda}]$, $b\in\bF$.
The value of $\bfC[\lambda][2i+b]$ corresponds to $\hat u^{(s_{[m-\lambda]})}_\vp$, where $b=\vp\bmod 2$ and $i=J(s_{[m-\lambda]})$.

\pairalg{
\caption{\texttt{CalcT\_SC}$(\vp)$}
\label{alg:calctsc}

\If{$\vp\! =\!n\!-\!2$}{\Return{$\Delta \mu_{1,1,1}(\bfT[m][0],\bfC[m][1])$}} \label{l:calctsc-n-2}
\If{$\vp =n-1$}{
    $a\gets \bfC[m-1][2]$, $b\gets \bfC[m][0]$\\
    \Return{$\Delta \mu_{2,1,0}\left(\bfT[m][0],a\oplus b,b\right)$\\ \label{l:calctsc-n-1}} 
}
$\lambda_* \gets  m$\\
\lIf*{$\vp =0$}{%
\label{l:calctsc-phi00} 
    \For{$\lambda \gets  2 \dots m-1$}{ 
      \For{$i \in [2^{m-\lambda}]$}{%
            $\bfT[\lambda][i]\! \gets\! \sigma_{0,2,2}(\bfT[\lambda\!-\!1][i], \bfT[\lambda\!-\!1][i\!+\!2^{m-\lambda}])$\label{l:calctsc-phi01}%
        }%
    }%
}\lElse*{\While{$\vp\text{ is odd} \wedge \lambda_*\!>\!2 \wedge \vp\!\neq\! 1$}{ \label{l:calctsc-lmin0}
        $\vp\gets \floor{\vp/2}, \; \lambda_\text{min}\gets \lambda_\text{min}-1$ \label{l:calctsc-lmin1}
    }
}
\For{$\lambda \gets  \lambda_* \dots m$}{ \label{l:calctsc-main0}
    $\sigma\gets $ operator for $\lambda,\vp$ in  Table~\ref{t:sfop} \label{l:calctsc-op0}\\
    \For{$i  \in [2^{m-\lambda}]$}{
        $\bfT[\lambda][i]\!\gets\!\sigma(\bfT[\lambda\!-\!1][i],\bfT[\lambda\!-\!1][i\!+\!2^{m-\lambda}], \bfC[\lambda][2i], \bfC[\lambda][2i\!\!+\!\!1])$ \label{l:calctsc-op1}
    }
    \lIf{$\vp\neq 0$}{$\vp \gets  2\vp+1$ \label{l:calctsc-main1}}
}
\Return{$\Delta \mu_{2,1,0}(\bfT[m][0])$}
}{
\caption{\texttt{UpdateC\_SC}$(\vp)$}
\label{alg:updatecsc}

$\lambda \gets  m, \; N\gets 1$ \\
\While{$\vp \neq 0 \wedge \; \lambda \neq 0$}{ 
  $C\gets\bfC[\lambda],\;D\gets\bfC[\lambda-1]$ \\
  $\psi \gets  \floor{\frac{\vp-1}{2}},\; b\gets \psi \bmod 2$\label{l:updatecsc-odd0}\\
  \If{$\vp \equiv 1 \bmod 2$}{ \label{l:updatecsc-com1}
    \For{$i \in[N]$}{
      \If{$\lambda = 1$}{
        $D[i]\gets C[2i]\oplus C[2i+1]$\\ \label{l:updatecsc-l10}
        $D[i+N]\gets C[2i+1]$} \label{l:updatecsc-l11}
      \Else{
        $D[2i+b]\gets C[2i]\oplus C[2i+1]$ \label{l:updatecsc-partsum0}\\
        $D[2i+b+2N]\gets C[2i+1] \label{l:updatecsc-partsum1}$
    }
    \lIf{$\vp \neq 2^\lambda\!-\!1$}{\textbf{break}}}\label{l:updatecsc-odd1}
  }
  \lElse*{\label{l:updatecsc-even0}
    \For{$i \in[N]$}{  
      $D[2i\!+\!b]\!\opgets\!C[2i]$; $D[2i\!+\!b\!+\!2N]\!\opgets\! C[2i]$ \label{l:updatecsc-even1}
    }
  }
  $\lambda \gets \lambda \!-\!1, \; \vp\gets \psi, \; N\gets 2N$ \label{l:updatecsc-com2}
}
}

\textbf{Alg.~\ref{alg:calctsc}} computes $\tL_\vp$, defined by \eqref{eq:tldef}, performing cluster operations defined by \eqref{eq:recl} and shown in Table~\ref{t:sfop}.
In lines~\ref{l:calctsc-n-2}--\ref{l:calctsc-n-1} the cases of $\vp\geq n-2$ are processed, following \eqref{eq:tllast}.
In lines~\ref{l:calctsc-phi00}--\ref{l:calctsc-phi01} the case of $\vp=0$ is processed.
On each layer except for the layer $m$, cluster operator $\sigma_{0,2,2}$ is called (see Fig.~\ref{fig:s022}) to obtain $2$-clusters which are needed initially.
When we access these layers for the second time, we will perform operator $\sigma_{0,3,1}$, as we will need $3$-clusters instead of $2$-clusters for a proper recursion \eqref{eq:recl}.
We need to compute new clusters on layer $\lambda-1$, when the phase on layer $\lambda$ is odd. 
Thus, for $\vp>0$, in lines~\ref{l:calctsc-lmin0}--\ref{l:calctsc-lmin1} the deepest cluster layer, which should be updated, is computed and stored in $\lambda_*$.
If the local phase on layer $\lambda$ is $\vp=2\psi+1$, we  proceed to the previous layer $\lambda-1$ and the local phase is now $\psi$.
The main loop in lines~\ref{l:calctsc-main0}--\ref{l:calctsc-main1} performs cluster operators on each layer.
The proper operator for local phase $\vp$ and layer $\lambda$ is chosen as given in Table~\ref{t:sfop}.
Two last hard decisions, stored in $\bfC[\lambda][2i]$ and $\bfC[\lambda][2i+1]$, are provided for the unified signature of operators, although in $\sigma_{i,t,j}$ for $i<2$
some of them are ignored.

\onetable{
\caption{Cluster operators on layer $2\leq \lambda \leq m$ in the SF SC decoding, not including operators $\Delta\mu$ for layer $m$.}
\label{t:sfop}
\centering
\begin{tabular}{|c|c|c|}
\hline
Phase & Operators & Complexity  \\\hline
$0$ & $\sigma_{0,2,2},\sigma_{0,3,1}$ & $52$ \\\hline
$1,3,...,2^\lambda-5$ & $\sigma_{1,3,2}$ & $\boldsymbol{56}$ \\\hline
$2,4,...,2^\lambda-4$ & $\sigma_{2,3,1}$ & $\boldsymbol{24}$ \\\hline
$2^\lambda-3$ & $\sigma_{1,3,0}$ & $8$ \\\hline
$2^\lambda-2$ & $\sigma_{2,2,0}$ & $4$ \\\hline
Total & & $40\cdot 2^\lambda\!-\!96$ \\\hline
\end{tabular}
}

\textbf{Alg.~\ref{alg:updatecsc}} propagates hard decisions $\hat u_{[\vp+1]}$ onto previous layers to obtain $\hat u^{(s)}_\psi$
by \eqref{eq:u01}.
Note that for recursion \eqref{eq:recl} we need at most two last hard decisions on $\hat u$.
The two last hard decisions, corresponding to last previous even and last previous odd phase on layer $\lambda$ for the $i$-th CvPT of size $2^\lambda$, are stored in $\bfC[\lambda][2i]$ and $\bfC[\lambda][2i+1]$, respectively.
When $\vp=2\psi+1$, we propagate partial sums as $\hat u^{(0)}\gets\hat u_{2\psi}\oplus \hat u_{2\psi+1}$ and $\hat u^{(1)}\gets\hat u_{2\psi+1}$, which is reflected in lines~\ref{l:updatecsc-odd0}--\ref{l:updatecsc-odd1}.
In lines~\ref{l:updatecsc-l10}--\ref{l:updatecsc-l11} we process layer $\lambda=1$, also performing a permutation to ensure the compatibility with the encoder.
When $\vp=2\psi+2$, we update partial sums as $\hat u^{(0)}_\psi\opgets\hat u_{2\psi+2}$ and $\hat u_\psi^{(1)}\opgets\hat u_{2\psi+2}$, which is reflected in lines~\ref{l:updatecsc-even0}--\ref{l:updatecsc-even1}.

\onealg{
\caption{$\call{Decode\_SC}(n, Y_{[n]}, \bfI)$}
\label{alg:decodesc}

allocate $\bfT[0..m]$, $m=\log_2n$\\
\For{$\lambda \in[m+1]$}{
    $\bfT[\lambda]\gets $ 2D array of size $2^{m-\lambda}\times8$
}
\lFor*{$i \in[n/2]$}{\label{l:decodesc-init0}
\For{$(a,b) \in \bF^2$}{$\bfT[1][i][a,b]\gets (a\oplus b)\cdot Y_i+b\cdot Y_{i+n/2}$}}\label{l:decodesc-init1}
\For{$\vp \gets  0 \dots n-1$}{
$l\gets \call{CalcT\_SC}(\vp,\bfT,\bfC)$ \label{l:decodesc-ct}\\
$\bfC[m][\vp \bmod 2]\!\gets\! \mathbf 1[\vp \in \mI \wedge l < 0]$ \\ \label{l:decodesc-hd}
$\call{UpdateC\_SC}(\vp,\bfC)$ \label{l:decodesc-uc}
}
$\hat u_{[n]}\gets \bfC[0][0..n-1](Q^{(n)})^{-1}$ \label{l:decodesc-cw2data}\\
\Return{$\hat u_{\mI}$} \label{l:decodesc-ret}
}

\textbf{In Alg.~\ref{alg:decodesc}} the top-level decoding function
is provided.
The inputs are the length of the code $n=2^m$, the channel LLRs $Y_{[n]}$, and the set of non-frozen positions $\mI\subseteq[n]$.
Array $Y$ of channel LLRs defined for channel output $y_{[n]}$ as 
\begin{align}
Y_i=\ln\frac{W(1|y_i)}{W(0|y_i)}.
\label{eq:Ydef}
\end{align}
The algorithm returns the estimated information vector $\hat u_{\mI}$.

By Remark~\ref{rm:clusteradd}, we can add an arbitrary constant value to all elements within a cluster.
In lines~\ref{l:decodesc-init0}--\ref{l:decodesc-init1} the values of
\begin{align*}
&L^{(s_{[m-1]})}_0[a,b]-\ln W(0|y_i)-\ln W(0|y_{i+n/2})
\splitwo
=\ln \frac{W(a\oplus b|y_i)}{W(0|y_i)}+\ln \frac{W(b|y_{i+n/2})}{W(0|y_{i+n/2})}
\splitone
=(a\oplus b)\cdot Y_i+b\cdot Y_i,
\end{align*}
are computed and stored in array $\bfT[1][i][a,b]$, where $i=J(s_{[m-1]})$.
Here, for $a\in\bF$, $r\in\bR$, product $a\cdot r\in\bR$ is computed as if $a$ was an integer.

In line~\ref{l:decodesc-ct}, function \call{CalcT\_SC} is called, which returns LLR $\tL_\vp$.
In line~\ref{l:decodesc-hd} the hard decision on $u_\vp$ is made by \eqref{eq:hd}.
In line~\ref{l:decodesc-uc} function \call{UpdateC\_SC} is called to propagate hard decisions $\hat u_\vp$, made on layer $m$, to previous layers.
In line~\ref{l:decodesc-cw2data} the estimated codeword is converted to the input information
vector by multiplying by $(Q^{(n)})^{-1}$, which can be done in $O(n\log n)$ operations by reversing the order of XOR nodes presented in Fig.~\ref{fig:cvpt}.
%
%
%
%


\subsection{Complexity of the Straightforward SC Decoding}
\label{s:cvpccomplsf}
We express the complexity in terms of the number of floating-point additions and comparisons, assuming that the complexity of bit operations is zero.
Also, we do not count comparison with zero and taking the absolute value as a comparison operation, since they can be implemented by taking/dropping the sign of a float, which can be considered as bit operations rather than float comparisons.




Operator $\sigma_{i,t,j}$ has complexity $\mC(\sigma_{i,t,j})=2^t\cdot (2^{j+1}-1)$, since it computes $2^j$ additions
and $2^j-1$ comparisons for each of $2^t$ output values.
Operator $\mu_{i,t,j}$ has complexity $\mC(\mu_{i,t,j})=2^t \cdot (2^j-1)$.
Operator $\Delta\mu_{i,1,j}$ has complexity $\mC(\Delta\mu_{i,1,j})=2\cdot(2^j-1)+1=2^{j+1}-1$.
%

The complexity of processing all phases on layer $\lambda=2..m-1$ is shown in Table~\ref{t:sfop}.
The total complexity of processing layer $\lambda$ is $2^{m-\lambda}\cdot(40\cdot 2^\lambda - 96)=40n-96n/2^\lambda$.
Layer $\lambda=1$ is processed in lines \ref{l:decodesc-init0}--\ref{l:decodesc-init1}, and its complexity is $\mC(n,1)=n/2$.
Compared to layers $2...m-1$, processing on layer $m$ includes converting the 3-cluster to 1-cluster by $\mu_{0,1,2},\mu_{1,1,1},\mu_{2,1,0}$ for phase $\vp<n-2$, $\vp=n-2$, $\vp=n-1$, respectively,
and then computing LLR by \eqref{eq:tldef}. 
The total complexity of all these operators is $7n-10$. 

The total complexity of cluster operators is
\begin{align*}
\mC(n)\!=\!7.5n\!-\!10\!+\!\sum_{\lambda=2}^m\! \left[40n\!-\!\frac{96n}{2^\lambda}\right]\!\!=\!40n\log_2 n\!-\!120.5n\!+\!86.
\end{align*} 
In the case of small $n$, the complexity of procesing the begining and ending phases $0$, $n-2$, $n-1$, and top/bottom layers $m$, $1$, $2$, can play a huge role in the overall complexity.
However, as $n\to\infty$, the complexity of computing output LLRs is only defined by two numbers, marked in bold in Table~\ref{t:sfop}. 
The term before $n\log_2 n$ is equal to the half-sum of these numbers.
These numbers correspond to the complexity of processing odd and even phases in most cases when $n\to \infty$.  


\section{Efficient SC Decoding \label{s:sceff}}

In this section we provide an SC decoder implementation which has  complexity $20n\log_2 n+o(n\log n)$ instead of $40n\log_2n+o(n\log n)$.
 
\subsection{Improved Sequence of Cluster Operators}

We start with a simple example which shows how one can reduce the complexity of cluster operators. 

Consider consecutive computing of $L_{2\psi+1}[u_{2\psi+1}^{2\psi+3}]$ and $L_{2\psi+2}[u_{2\psi+2}^{2\psi+4}]$ on some layer $3\leq\lambda\leq m-2$.
In the straightforward implementation, computing the former cluster costs $56$ operations, and computing the latter cluster costs $24$ operations.
Note that both $3$-clusters can be obtained by marginalization~\eqref{eq:mudef} of $4$-cluster $L_{2\psi+1}[u_{2\psi+1}^{2\psi+4}]$:
\begin{align}
L_{2\psi+1}[u_{2\psi+1}^{2\psi+3}]&=\mu_{0,3,1}(L_{2\psi+1}[u_{2\psi+1}^{2\psi+4}]) \label{eq:marginodd}\\
L_{2\psi+2}[u_{2\psi+2}^{2\psi+4}]&=\mu_{1,3,0}(L_{2\psi+1}[u_{2\psi+1}^{2\psi+4}],\hat u_{2\psi+1}) \label{eq:margineven}
\end{align}
A $4$-cluster $L_{2\psi+1}[u_{2\psi+1}^{2\psi+4}]$ can be computed as
\begin{align}
&L_{2\psi+1}[x_1^4]=_{x_0=\hat u_{2\psi}}\max_{x_5\in\bF}\set{L^{(0)}_{\psi}[x_{[6]}X^{(6)}]+L_\psi^{(1)}[x_{[6]}Z^{(6)}]}
\label{eq:sigma141}
\end{align}
which requires $2$ additions and a comparison for each of $16$ output values, making the total complexity $48$ operations.
After that, we perform marginalization~\eqref{eq:marginodd}--\eqref{eq:margineven}.
Marginalization operator~\eqref{eq:marginodd} has complexity $8$ as it performs  a comparison
for each of the $8$ values of $L_{2\psi+1}[u_{2\psi+1}^{2\psi+3}]$.
The second operator~\eqref{eq:margineven} has zero complexity, as it consists of assigning some values of the $4$-cluster to $3$-cluster $L_{2\psi+2}[u_{2\psi+2}^{2\psi+4}]$.

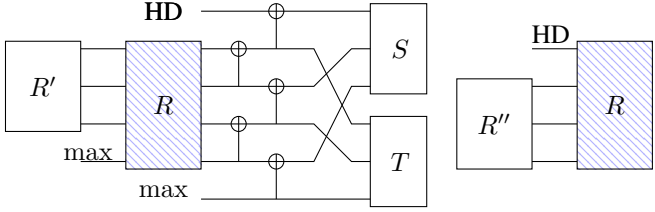
\begin{figure}
\centering
\newcommand{\opwidth}{0.23\textwidth}
\newcommand{\opx}{0.5cm}
\newcommand{\opy}{0.5cm}
\newcommand{\rightx}{15.5}
\begin{tikzpicture}[x=\opx,y=\opy]
\def\r{0.1cm}
\draw (11,6)--(\rightx,6);
\draw (11,5)--(14,5)--(15,3)--(\rightx,3);
\draw (11,4)--(14,4)--(15,5)--(\rightx,5);
\draw (11,3)--(14,3)--(15,2)--(\rightx,2);
\draw (11,2)--(14,2)--(15,4)--(\rightx,4);
\draw (11,1)--(\rightx,1);

\draw (12,5) circle(\r);
\draw (12,3) circle(\r);
\draw (13,6) circle(\r);
\draw (13,4) circle(\r);
\draw (13,2) circle(\r);

\draw (12,2)--(12,3.2);
\draw (12,4)--(12,5.2);
\draw (13,1)--(13,2.2);
\draw (13,3)--(13,4.2);
\draw (13,5)--(13,6.2);
\draw (9,1.8)[pattern=north west lines, pattern color=blue!30!] rectangle (11,5.2);
\draw (\rightx,3.8) rectangle (\rightx+1.5,6.2);
\draw (\rightx,0.8) rectangle (\rightx+1.5,3.2);
\node at (10,1.2) {$\max$};
\node at (10,3.5) {$R$};
\node at (6.8,4) {$R'$};
\node at (10,6) {HD};

\foreach \y in {2,3,4,5}{
    \draw (7.8,\y)--(9,\y);
}

\node at (8,2.2) {$\max$};
\draw (5.8,2.8) rectangle (7.8,5.2);
\node at (10,6) {HD};
\def\sx{18.0}
\def\sy{0.0}
\draw (3+\sx,1.8+\sy)[pattern=north west lines, pattern color=blue!30!] rectangle (5+\sx,5.2+\sy);
\foreach \y in {2,3,4,5}{
    \draw (1.8+\sx,\y+\sy)--(3+\sx,\y+\sy);
}
\draw (-0.2+\sx,1.8+\sy) rectangle (1.8+\sx,4.2+\sy);
\node at (18.8,3) {$R''$};
\node at (16.3,2) {$T$};
\node at (16.3,5) {$S$};
\node at (22,3.5) {$R$};
\node at (2.3+\sx,5.4+\sy) {HD};

\end{tikzpicture}
\caption{Given $S=L_\psi^{(s,0)}(x_{[3]})$
and $T=L_\psi^{(s,1)}(x_{[3]})$, obtain a $4$-cluster $R=L_{2\psi+1}^{(s)}(x_{[4]})$, and use it to obtain $3$-clusters $R'=L_{2\psi+1}^{(s)}(x_{[3]})$
and $R''=L_{2\psi+2}^{(s)}(x_{[3]})$ by marginalization (compare to Fig.~\ref{fig:s132}--\ref{fig:s231}).}
\label{fig:cvpcpat-oddeff}
\end{figure}

The proposed approach is illustrated in Fig.~\ref{fig:cvpcpat-oddeff}.

Thus, instead of $80$ operations, one can compute two $3$-clusters in $48+8=56$ operations.
However, when asked for a $3$-cluster on an odd phase, layer $\lambda$ should save the computed $4$-cluster in some array for reuse on the subsequent even phase, which is solved by appropriate
tuning of Tal-Vardy data structures.

Similar tricks can be performed for other cases.
The complete description of the calling sequence of cluster operators $\sigma$ and marginalization operators $\mu$ are shown in Tables~\ref{t:op2}--\ref{t:opm}.

\subsection{Efficient Computing of Two Maxima}
\label{s:cvpcsmartmax}
Consider a problem of computing $R_{[2]}$ for input $S_{[2]}$ and $T_{[2]}$, such that
\begin{align}
R_0=\max\set{S_0\!+\!T_0,S_1\!+\!T_1}, 
R_1=\max\set{S_0\!+\!T_1,S_1\!+\!T_0} \label{eq:smdef}
\end{align}
The complexity of straightforward computation is $6$ operations.
However, one can do better.

%

Since in r.h.s. of \eqref{eq:smdef} we have all $4$ pairwise sums of $S_{[2]}$ and $T_{[2]}$, the maximum of the output values $R_r=\max\set{R_0,R_1}=\max\set{S_0,S_1}+\max\set{T_0,T_1}$.
Denote maxima of each input arrays as $S_s$ and $T_t$.
Then, $R_r=S_s+T_t$, and obviously $r=s\oplus t$.
The other output value, $R_{r\oplus 1}=R_{s\oplus t\oplus 1}$, can be expressed as 
\begin{align}
&R_{r\oplus 1}=\max\set{S_s+T_{t\oplus 1},S_{s\oplus 1}+T_t}
\splitwo
=\max\set{R_{r}-T_t+T_{t\oplus 1},R_{r}-S_s+S_{s\oplus 1}}
\nonumber\\&
=R_{r}-\min\set{T_t-T_{t\oplus 1},S_s-S_{s\oplus 1}}
\splitwo
=R_{r}-\min\set{|T_0-T_1|,|S_0-S_1|}.
\label{eq:smartmaxr1}
\end{align}
The last equality follows from the fact that $T_t\geq T_{t\oplus 1}$ and $S_s\geq S_{s\oplus 1}$.
So, instead of comparing $S_0$ with $S_1$ and $T_0$ with $T_1$ for computing $R_{s\oplus t}$, one can begin with computing $\delta=S_0-S_1$ and $\Delta=T_0-T_1$ and take their
signs for free.
Basing on their signs, one computes $R_r=S_s+T_t$. 
Also, the absolute values of $\delta$ and $\Delta$ are then used in~\eqref{eq:smartmaxr1}, i.e. $R_{r\oplus 1}=R_r-\min\set{|\delta|,|\Delta|}$.
The total complexity of such computation is $5$.

\onealg{
\caption{$\call{Max2D}(S_{[2]},T_{[2]},\delta,\Delta)$}
\label{alg:max2delta}

$s\gets \boldsymbol{1}[\sgn \delta = -1]$; $t\gets \boldsymbol{1}[\sgn \Delta = -1]$ \label{l:max2d-st}\\
$R_{s\oplus t}\gets S_s+T_t$; $R_{s\oplus t\oplus 1}\gets R_{s\oplus t}-\min\set{|\delta|,|\Delta|}$ \label{l:max2d-r01}
\Return{$R_{[2]}$}
}

Function \call{Max2D} in Alg.~\ref{alg:max2delta} uses already computed values of $\delta$ and $\Delta$ and computes $R_0^1$.
Its complexity is $3$ operations.

\subsection{Employing Efficient Maxima in Cluster Operators}
Consider cluster operator $\sigma_{1,4,1}$ given by \eqref{eq:sigma141}.
Assume that $\hat u_{2\psi}=0$ (otherwise we can permute the input), and denote $S=L^{(0)}_\psi$, $T=L^{(1)}_\psi$, $R=L_{2\psi+1}$.
Then, one obtains
\begin{align}
R[x_{[4]}]=\max\set{S[A_0(x)]\!+\!T[B_0(x)],S[A_1(x)]\!+\!T[B_1(x)]},
\label{eq:rx}
\end{align}
where 
\begin{align*}
A_b(x)&=(0,x_{[4]},b)X^{(6)}=(x_0\oplus x_1,x_1\oplus x_2\oplus x_3,x_3\oplus b)\\
B_b(x)&=(0,x_{[4]},b)Z^{(6)}=(x_0\oplus x_1,x_2\oplus x_3,b).
\end{align*}

Denote $x'_{[4]}=(x_0,x_1,x_2\oplus 1,x_3\oplus 1)$.
Observe that $A_b(x')=A_{b\oplus 1}(x)$, $B_b(x')=B_{b \oplus 1}(x)$ for all $b\in\bF$.
Thus, 
$
R[x'_{[4]}]=\max\set{S[A_1(x)]\!+\!T[B_0(x)],S[A_0(x)]\!+\!T[B_1(x)]}.
$
Comparing this with \eqref{eq:rx}, one can see that pair $(R[x], R[x'])$ can be obtained via function \call{Max2D}, with
$\delta=S[A_0(x)]-S[A_1(x)]$ and $\Delta=T[B_0(x)]-T[B_1(x)]$.

The output array $R$ consists of $8$ pairs $(R[x_{[4]}], R[x_{[4]}\oplus(0011)])$, $x_3=0$, $x_{[3]}\in\bF^3$.
For each of these pairs $\delta=S[A_0(x)]-S[A_1(x)]$. 
Furthermore, it appears that we have only $4$ different values of $\delta$.
Indeed, consider $\ox=x+(1,1,1,0)$.
Note that $A_b(\ox)=A_b(x)=A_{b\oplus 1}(\ox')=A_{b\oplus 1}(x')$.
So, for two pairs of output values, $(R[x],R[x'])$ and $(R[\ox,\ox'])$ we have the same value of $\delta$.
The same applies to $\Delta$ and pairs $(R[x],R[x'])$ and $(R[\tx,\tx'])$, where $\tx=x+(1,1,0,0)$.
We propose to compute $4$ values of $\delta$ and $4$ values of  $\Delta$, and then call function \call{Max2D} for $8$ times, see Alg.~\ref{alg:s141}.
This trick works for operator $\sigma_{0,3,1}$ as well (see Alg.~\ref{alg:s031}).

The complexity of the efficient implementation of operator $\sigma_{1,4,1}$ is $32$ (instead of $48$), the complexity of operator $\sigma_{0,3,1}$ is $16$ (instead of $24$).
\pairalg{
\caption{$\sigma_{1,4,1}(S,T,u)$, complexity: $32$}
\label{alg:s141}
allocate $4$-cluster $R$\\
\For{$(a,b)\in\bF^2$}{
    $\delta[a,b]\gets S[a,b,0]-S[a,b,1]$\\
    $\Delta[a,b]\gets T[a,b,0]-T[a,b,1]$\\
}
\For{$x \in\bF^3$}{
    $s\gets (u_0,x)X^{(4)}; t\gets (u_0,x)Z^{(4)}$\\
    $(R[x,0],R[x_{[2]},x_{2}\oplus 1,1])\gets\call{Max2D}(S[s,0..1],T[t,0..1],$ $\delta[x_{[2]}],\Delta[x_{[2]}])$
}
\Return{$R$}
}{
\caption{$\sigma_{0,3,1}(S,T,u)$, complexity: $16$ }
\label{alg:s031}

allocate $3$-cluster $R$\\
\For{$a\in\bF$}{
    $\delta[a]\gets S[a,0]-S[a,1]$\\
    $\Delta[a]\gets T[a,0]-T[a,1]$
}
\For{$(a,b) \in\bF^2$}{
    $(R[a,b,0],R[a,b\oplus 1,1])\gets\call{Max2D}(S[a\oplus b,0..1],T[b,0..1],$ $\delta[a\oplus b], \Delta[b])$
}
\Return{$R$}
}

\subsection{Processing of Layer $2$}
\label{s:cvpcl2eff}
\begin{figure}
\newcommand{\opwidth}{0.23\textwidth}
\newcommand{\opx}{0.4cm}
\newcommand{\opy}{0.5cm}
\newcommand{\rightx}{15.5}
\centering
\begin{tikzpicture}[x=\opx,y=\opy]
\def\r{0.1cm}
\foreach \y in {2,3,4}{
    \draw (15,\y)--(\rightx,\y);
}
\draw (11,4)--(\rightx+2,4);
\draw (11,3)--(14,3)--(15,2)--(\rightx,2)--(\rightx+2,2);
\draw (11,2)--(14,2)--(15,3)--(\rightx,3)--(\rightx+2,3);
\draw (11,1)--(\rightx+2,1);
\def\sy{0}
\def\sx{0}
\foreach \y in {1,2,...,4}{
    \draw (8.5,\y)--(9.7,\y);
}
\foreach \y in {2,3,4}{
    \draw (6,\y)--(7.2,\y);
}
\draw (7.2+\sx,1.8+\sy) rectangle (8.5,4.2+\sy);
\draw (9.7,0.8+\sy)[pattern=north west lines, pattern color=blue] rectangle (11,4.2);
\draw (4.7,2.8+\sy) rectangle (6,4.2+\sy);
\draw (12,3) circle(\r);
\draw (12,2)--(12,3.2);
\draw (13,2) circle(\r);
\draw (13,3)--(13,4.2);
\draw (13,4) circle(\r);
\draw (13,1)--(13,2.2);
\draw (16,3)--(16,4.2);
\draw (16,1)--(16,2.2);
\draw (16,2) circle(\r);
\draw (16,4) circle(\r);
\node at (\rightx+2.5,4.4) {$Y_{i}$};
\node at (\rightx+2.5,2.4) {$Y_{i+n/4}$};
\node at (\rightx+2.5,3.4) {$Y_{i+n/2}$};
\node at (\rightx+2.5,1.4) {$Y_{i+3n/4}$};
\node at (8.8,1.4) {$\max$};
\node at (6.3,2.4) {$\max$};
\def\sx{10.2}
\def\sy{4.2}
\foreach \y in {-0.2,-1.2,-2.2,-3.2}{
    \draw (14+\sx,\y+\sy)--(15.2+\sx,\y+\sy);
}
\foreach \y in {-1.2,-2.2,-3.2}{
    \draw (11.5+\sx,\y+\sy)--(12.7+\sx,\y+\sy);
}
\draw (15.2+\sx,-3.4+\sy)[pattern=north west lines, pattern color=blue] rectangle (16.5+\sx,0+\sy);
\draw (12.7+\sx,-3.4+\sy) rectangle (14+\sx,-1+\sy);
\draw (10.2+\sx,-3.4+\sy) rectangle (11.5+\sx,-2+\sy);
\node at (14.5+\sx,0.3+\sy) {$\hat u^{(s)}_0$};
\node at (12+\sx,-0.7+\sy) {$\hat u^{(s)}_1$};
\end{tikzpicture}
\caption{Efficient processing of layer $\lambda=2$: one operator $\sigma_{0,4,0}$
to obtain $4$-cluster $L^{(s)}_0(u^{(s)}_{[4]})$, and $4$ marginalization
operators applied to the $4$-cluster to obtain the remaining clusters
(compare to Fig.~\ref{fig:s022}--\ref{fig:s220}).}
\label{fig:cvpcpat-l2eff}
\end{figure}
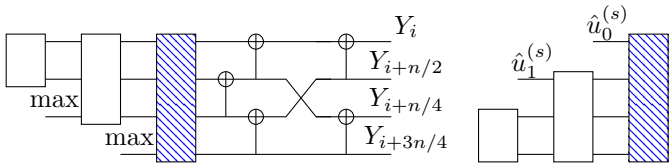

In order to enable computations on layer $3$, layer $2$ should provide clusters $L^{(s)}_0[v_{[3]}]$, $L^{(s)}_0[v_{[2]}]$ and $L_{2,1}[v_1^3]$, where $v_i\in\bF$, $s\in\bF^{m-2}$.
We propose to compute a $4$-cluster $L^{(s)}_0[v_{[4]}]$ and obtain all needed clusters by calling appropriate marginalization
operators.
The proposed approach is illustrated in Fig.~\ref{fig:cvpcpat-l2eff}.
Denote 
$
x_{[4]}\!=\!v_{[4]}Q^{(4)}
$
.
By Remark~\ref{rm:clusteradd}, we can add arbitrary value to all cluster elements.
Therefore, we can compute $L^{(s)}_0[v_{[4]}]$ as
\begin{align}
&\sum_{i=0}^{3}\ln W(x_i|y_{j+in/4})\!+\!\frac{1}{2}\sum_{i=0}^{3}\ln\left[W(0|y_{j+in/4}) W(1|y_{j+in/4})\right]
\splitwo
=\sum_{i=0}^3(-1)^{x_i}\cdot\frac{Y_{j+in/4}}{2},
\label{eq:llr2}
\end{align}
where $j=J(s)$ given by \eqref{eq:jdef}, $Y_i$ is given by \eqref{eq:Ydef}.

The r.h.s. of \eqref{eq:llr2} for all $x_{[4]}\in\bF^4$ can be computed as follows.
First, arrange all $x_{[4]}\in\bF^4$ in the order of Gray code as $(x^{(i)})_{i=0}^{15}$, $x^{(i)}\in\bF^4$.
Arrange\footnote{A fun fact: $v^{(i)}$ is then a binary representation of $i$.} all $v^{(i)}_{[4]}\in\bF^4$ according to the order of $x^{(i)}$, so that $x^{(i)}=v^{(i)}Q^{(4)}$.
Consider the first $8$ values.
They correspond to $x_3=0$, which implies $v_3=0$.
Compute $L^{(s)}_0[v^{(0)}]=L^{(s)}_0[\bnull^4]=\frac{1}{2}\sum_{i=0}^3Y_{j+in/4}$ by three additions (we do not count dividing
by $2$ as a floating-point operation, since it can be easily implemented by a binary shift).
Each of the next $7$ values of $L^{(s)}_0[v^{(h)}]$, $h=1..7$ can be computed with complexity $1$ by adding or subtracting $Y_{j+in/4}$ from $L_0^{(s)}[v^{(h-1)}]$, where $i$ is the index where $x^{(h-1)}$ and $x^{(h)}$ differ.
Note also that $(v_{[3]},0)Q^{(4)}=(v_{[3]},1)Q^{(4)}\oplus (1,1,1,1)$, so the remaining $8$ values can be computed as 
$L^{(s)}_0[v_{[3]},1]=-L^{(s)}_0[v_{[3]},0]$ with complexity $0$.
Moreover, $$L^{(s)}_0[v_{[3]}]=\max\set{L^{(s)}_0[v_{[3]},0],L^{(s)}_0[v_{[3]},1]}=|L^{(s)}_0[v_{[3]},0]|$$ can be computed with zero complexity as well.

In the case when the half of the cluster $\sigma_{i,t,j}(A,B)$ is equal to the other half of the cluster with an opposite sign, we denote such cluster operator by $\osigma_{i,t,j}$ meaning the operator with reduced complexity.
This operator is given by
\begin{align}
&\osigma_{i,t,0}(A,B,u)[x_{[t-1]},0]=-\osigma_{i,t,0}(A,B,u)[x_{[t-1]},1]=
\splitwo
A[(u_{[i]}, x_{[t-1]}, 0)X^{(2t')}]+B[(u_{[i]}, x_{[t-1]}, 0)Z^{(2t')}].
\label{eq:osigmadef}
\end{align}
The same is applied to operator $\omu_{i,t,j}$, given by
\begin{align}
\omu_{i,t,1}(A,u)[x_{[t]}]=|A[u_{[i]},x_{[t]},0]|
\label{eq:omudef}
\end{align}

%


\subsection{Complexity of the Efficient Implementation}
\onetable{
\caption{Cluster operators on layer $2$.}
\label{t:op2}
\begin{tabular}{|c|c|c|}
\hline
Phases & Cluster Operators & Complexity\\\hline
$0$ & $\osigma_{0,4,0}, \omu_{0,3,1},\mu_{0,2,1}$ & $14$ \\\hline
$1$ & $\mu_{1,3,0}$ & $0$ \\\hline
$2$ & $\mu_{1,2,0}$ & $0$ \\\hline
Total & & $14$\\\hline
\end{tabular}
}

\pairtable{
\caption{Cluster operators on layer $\lambda$, $3\leq \lambda \leq m-2$.}
\label{t:op3m2}
\begin{tabular}{|c|c|c|}
\hline
Phases & Cluster Operators & Complexity\\\hline
$0$ & $\sigma_{0,3,1}, \mu_{0,2,1}$ & $20$ \\\hline
$1,3,...,2^{\lambda}-7$ & $\sigma_{1,4,1},\mu_{0,3,1}$ & $\boldsymbol{40}$ \\\hline
$2,4,...,2^{\lambda}-6$ & $\mu_{1,3,0}$ & $\boldsymbol{0}$ \\\hline
$2^{\lambda}-5$ & $\osigma_{1,5,0},\omu_{0,4,1},\mu_{0,3,1}$ & $24$ \\\hline
$2^{\lambda}-4$ & $\mu_{1,3,0}, \mu_{1,4,0}$ & 0 \\\hline
$2^{\lambda}-3$ & $\mu_{1,3,0}$ & 0 \\\hline
$2^{\lambda}-2$ & $\mu_{1,2,0}$ & 0 \\\hline
Total & & $20\cdot 2^\lambda-76$\\\hline
\end{tabular}
}{
\caption{Cluster operators on  layer $m-1$.}
\label{t:opm1}
\begin{tabular}{|c|c|c|}
\hline
Phases & Cluster Operators & Complexity\\\hline
$0$ & $\sigma_{0,3,1}, \mu_{0,2,1}$ & $20$ \\\hline
$1,3,...,\frac{n}{2}\!-\!5,\frac{n}{2}\!-\!2$ & $\mu_{1,2,0}$ & $\boldsymbol{0}$ \\\hline
$2,4,...,n/2-6$ & $\sigma_{2,3,1}, \mu_{1,2,1}$ & $\boldsymbol{24}$ \\\hline
$n/2-4$ & $\osigma_{2,4,0}, \omu_{0,3,1}, \mu_{0,2,1}$ & $12$ \\\hline
$n/2-3$ & $\mu_{1,2,0}, \mu_{1,3,0}$ & $0$ \\\hline
Total & & $6n-40$\\\hline
\end{tabular}
}

\pairtable{
\caption{Cluster operators on  layer $m$.}
\label{t:opm}
\begin{tabular}{|c|c|c|}
\hline
Phases & Cluster Operators & Complexity\\\hline
$0$ & $\sigma_{0,3,1}, \mu_{0,2,1}, \Delta\mu_{0,1,1}$ & $23$ \\\hline
$1,n-2$ & $\Delta\mu_{1,1,0},\mu_{1,2,0}$ & $1$ \\\hline
$2,n-1$ & $\Delta\mu_{1,1,0}$ & $1$ \\\hline
$3,5,...,n-5$ & $\sigma_{1,2,1},\Delta\mu_{0,1,1}$ & $\boldsymbol{13}$ \\\hline
$4,6,...,n-4$ & $\Delta\mu_{1,1,0}$ & $\boldsymbol{1}$ \\\hline
$n-3$ & $\osigma_{1,3,0},\omu_{0,2,1}, \Delta\mu_{0,1,1}$ & $7$ \\\hline
Total & & $7n-8$\\\hline
\end{tabular}
}{
\caption{Complexity of computing output LLRs for $Q^{(n)}$.}
\label{t:optotal}
\centering
\begin{tabular}{|c|c|c|c|c|}
\hline
$n$ & SF & Proposed &  Proposed/Arikan \cite{arikan2009channel}\\\hline
$16$ & $718$ & $272$ &  $4.25$ \\\hline
$32$ & $2630$ & $968$ &  $6.05$ \\\hline
$64$ & $7734$ & $3000$ &  $7.81$ \\\hline
$128$ & $20502$ & $8344$  & $9.31$ \\\hline
$1024$ & $2.86\cdot 10^5$ & $1.27\cdot 10^5$ & $12.37$ \\\hline
$4096$ & $1.47\cdot 10^6$ & $6.70\cdot 10^5$  & $13.63$ \\\hline
$16384$ & $7.20\cdot 10^7$ & $3.33\cdot 10^7$  & $14.54$ \\\hline
\end{tabular}
}

The complexity of efficient operator $\sigma_{i,t,1}$, employing function \call{Max2D}, is $5\cdot 2^{t-1}$, with the exception for $\sigma_{1,4,1}$ and $\sigma_{1,3,1}$, where we reuse some values of $\delta, \Delta$, see Alg.~\ref{alg:s141}--\ref{alg:s031}.
The complexity of operators $\mu_{i,t,j}$ and $\Delta\mu_{i,1,j}$ is $2^t\cdot(2^j-1)$ and $2\cdot(2^j-1)+1$, respectively.
The complexity of operator $\omu$ is $0$.
The complexity of operator $\osigma_{i,t,0}$ is $2^{t-1}$.

The total complexity $\omC(n)$ of cluster operators in the efficient implementation of SC decoding can be computed from Tables~\ref{t:op2}--\ref{t:opm} as $\omC(n)=\sum_{\lambda=2}^m\omC(n,\lambda)$,
where $\omC(n,\lambda)$ is the complexity of processing layer $\lambda$ and is given by 
\begin{align}
&\omC(n)=3.5n+(12n\!-\!80)+(7n\!-\!8)+\!\!\!\sum_{\lambda=3}^{\log_2n-2}\left(20n-\frac{76n}{2^{\lambda}}\right)
\splitwo
=20n\log n-76.5n+216.
\end{align}

In Table~\ref{t:optotal} the complexity of the proposed LLR computation is compared to that of the straightforward
LLR computation.
Also there is a relative complexity of proposed algorithm of processing $Q^{(n)}$ to the complexity of computing LLRs at
the output of Arikan polarizing transformation of size $n$, which is $n\log_2 n$.
With $n\to\infty$, this relation tends to $20$, since the complexity of computing LLRs for $Q^{(n)}$ is $20n\log n+o(n \log n)$.
However, one can see that for practical code lengths the relation is much smaller than $20$.


\section{Efficient List Decoding Implementation}
\label{s:listeff}

The straightforward extension of SC decoder from Section~\ref{s:sceff} to the case of list decoder leads to copying vast amount of clusters, especially in the case of large $l$.
In this section we propose an efficient list decoder, which does not copy any clusters.

\subsection{General Structure of the List Decoding Algorithm}
We employ Tal-Vardy \cite{tal2015list} approach to list decoding.
At each phase $\vp$, we have a list of $l'$ paths $x_{[\vp]}$, $i\in[l']$ together with their scores $S(x_{[\vp]})$, defined as
\begin{align}
S(x_{[\vp]})=\ln \max_{x_{\vp}^{n-1}\in\bF^{n-\vp}} W^n(x_{[n]}Q^{(n)}|y_{[n]}).
\label{eq:sdef}
\end{align}
If $\vp\in\mI$, each path $x_{[\vp]}$ has two possible continuations $x_{[\vp+1]}$, corresponding to two possible values of $x_{\vp}\in\bF$.
If $\vp\in\mF$, each path $x_{[\vp]}$ has only one possible continuation $x_{[\vp+1]}$ with $x_{\vp}=0$.
At phase $\vp$ we consider all possible continuations of all $L'$ paths.
So, there can be $l'$ or $2l'$ continuations, depending on whether $\vp\in\mI$ or not.
The maximum list size is restricted by manually chosen parameter $l$, which allows balancing between ML decoding and SC decoding.
When $\vp\in\mI$, and $2l'>l$, we compute score of all $2l'$ paths and throw away $2l'-l$ paths that have the least score.
It can be shown that
\begin{align}
S(x_{[\vp+1]})=S(x_{[\vp]})+\tau(x_{\vp}, \tL_\vp),
\label{eq:srec}
\end{align}
where
\begin{align}
\tau(x_\vp,\tL_\vp)=\begin{cases}0, & \text{if }(-1)^{x_\vp}=\sgn \tL_\vp\\|\tL_\vp| & \text{otherwise}\end{cases}
\label{eq:taudef}
\end{align}
Thus, at each phase $\vp$  we should compute $\tL_\vp$ for each path in the list.
Note that the score of the empty path $\ve$ is 
\begin{align}
S(\ve)=\ln \max_{x_{[n]}\in\bF^{n}} W^n(x_{[n]}Q^{(n)}|y_{[n]})=\sum_{i\in[n]}\max_{a\in\bF}\ln W(a|y_i),
\label{eq:emptyscore}
\end{align}
since $Q^{(n)}$ is invertible.
For a given input vector $y_{[n]}$, this number is the same additive term for the score of all paths, so we can set $S(\ve)=0$
without influencing the behaviour of the list decoder.

\subsection{List Decoding Implementation}
In Alg.~\ref{alg:decode}--\ref{alg:clonepath} the efficient list decoding implementation is presented.

\pairalg{
\caption{\call{Decode}$(n, Y_{[n]}, \mI, l)$}
\label{alg:decode}
\KwOut{Estimated data vector}

$p_0\gets\call{Init}()$\\
$T\gets \call{GetPtr\_T}(p_0,2,4)$\\ \label{l:loadllr-gett4}
\For{$i \in[n/4]$}{ \label{l:loadllr-l20}
    $T[i]\gets\osigma_{0,4,0}(Y[i,i+\frac{n}{4},i+\frac{n}{2},i+\frac{3n}{4}])$\\
} \label{l:loadllr-l21}
\For{$\vp \gets  0 \dots n-1$}{ \label{l:decode-main0}
    $\call{CalcT}(p,\vp)$\\
    \lIf{$\vp\in\mI$}{$\call{ContinueInfo}(\vp)$}\lElse{$\call{ContinueFrozen}(\vp)$}
    $\call{UpdateC}(\vp)$
} \label{l:decode-main1}
$p\gets\arg\max_{i\in\bfA}\var{Score}[i]$ \label{l:decode-best}\\
%
$C\gets\call{CGetPtr\_C}(p,0)$\\
$\hat u_{[n]}\gets C[0..n-1](Q^{(n)})^{-1}$ \\
\Return{$\hat u_{\mI}$} \label{s:l-ret}
}{
\caption{\texttt{Init}$()$}
\label{alg:initlist}

\KwOut{the index of the initial empty path}
\For{$(p,\lambda)\in[l]\times[m+1]$}{
    $\bfC[\lambda][p]\gets $ bit array of size $2^{m-\lambda+1}$\\
}
\For{$(p,\lambda,d)\in[l]\times[m+1]\times[4]$}{
    $\bfT[\lambda][p][d]\gets $ 2D real array of size $2^\lambda \times 2^{d+2}$
}    
$\var{FreeArrayIndices}[0..m][0..4]\gets$ stacks with set $[l]$ \\
$\var{RefCount}[0..m][0..l-1][0..4]\gets 0$\\
$\var{RefCount}[0..m][l][0..4]\gets +\infty$\\
$\var{ArrayIndex}[0..m][0..l-1][0..4]\gets -1$\\
$\var{ArrayIndex}[0..m][0][0..4]\gets l$\\
$\var{Score}[0..l-1]\gets 0$\\
$\bfA\gets\set{0}$;
$\bfR\gets $ real array of size $l$\\
\Return{$0$}
}

In  \textbf{Alg.~\ref{alg:decode}}, the top-level function \call{Decode} calls function \call{Init} to allocate and initialize data structures.
\call{ContinueFrozen} and \call{ContinueInfo} for processing frozen and information symbols.

In \textbf{Alg.~\ref{alg:initlist}}, function \call{Init} initializes data structures.
In the efficient implementation, we use $D$-clusters for $D\in\set{2,3,4,5}$ on each layer.
They are modified independently, so to avoid copying of clusters we propose to keep array indices and reference counters for each cluster dimension separately.
This leads to extra dimension of arrays $\bfT$, \var{ArrayIndex}, \var{FreeArrayIndices}, \var{RefCount}.

Array $\bfT$ is indexed as $\bfT[\lambda][p][D][i][x_{[D+2]}]$, where $\lambda\in[m]\setminus[2]$ is the layer, $p\in[l]$ is the path index, $D\in[4]$ is the cluster dimension
minus $2$, $i\in[2^{m-\lambda}]$ is the index of CvPT $Q^{(2^\lambda)}$ in the recursive expansion \eqref{eq:cvpt}, which corresponds to $u^{(s)}$ and $y^{(s)}$ for $i=J(s)$.
Finally, $x_{[D+2]}\in\bF^{D+2}$ is the index within the cluster, this index corresponds to the value of not-yet-estimated input symbols $u_{\vp..\vp+D+1}^{(s)}$.

Other variables are indexed as $\var{ArrayIndex}[\lambda][p][d]$, $\var{FreeArrayIndices}[\lambda][d]$, and $\var{RefCount}[\lambda][p][d]$, where $\lambda$, $p$ are layer
and path index, the  value of $d=0$ corresponds to array indices for $\bfC$ array, the values of $d=1,...,4$ correspond to array indices for $(d+1)$-clusters.

%
%
%
%
%

\pairalg{
\caption{\call{GetPtr\_T}$(p,\lambda, D)$}
\label{alg:getpointert}

$d\gets D-1$;
$i\gets \var{ArrayIndex}[\lambda][p][d]$ \\
\lIf*{$\var{RefCount}[\lambda][i][d]\!\leq\! 1$}{\Return{$\bfT[\lambda][i][d\!-\!1]$}}\\ 
$\var{RefCount}[\lambda][i][d]\mgets1$ \\
$j\gets \call{Pop}(\var{FreeArrayIndices}[\lambda][d-1])$\\
$\var{RefCount}[\lambda][j][d]\gets 1$;
$\var{ArrayIndex}[\lambda][p][d]\gets j$ \\
\Return{$\bfT[\lambda][j][d-1]$}
}
{
\caption{\call{GetPtr\_C}$(p,\lambda)$}
\label{alg:getpointerc}

$i\gets \var{ArrayIndex}[\lambda][p][0]$\\
\lIf*{$\var{RefCount}[\lambda][i][0]\leq 1$}{%
    \Return{$\bfC[\lambda][i]$} 
}
$\var{RefCount}[\lambda][i][0]\mgets 1$ \\
$j\gets \call{Pop}(\var{FreeArrayIndices}[\lambda][0])$ \\
$\var{ArrayIndex}[\lambda][p][0]\gets j$ \\
$\var{RefCount}[\lambda][j][0]\gets 1$ \\
$\bfC[\lambda][j][0..2^{m-\lambda+1}-1] \gets \bfC[\lambda][i][0..2^{m-\lambda+1}-1]$ \label{l:getpointerc-copyc}\\
\Return{$\bfC[\lambda][j]$}
}

\textbf{Alg.~\ref{alg:getpointert}}, function \call{GetPtr\_T} returns a pointer to a free array of clusters of given dimension for writing.
When updating clusters, we do not need their old values, so we do not copy subarrays of $\bfT$.

\textbf{Alg.~\ref{alg:getpointerc}}, function \call{GetPtr\_C} returns a pointer to a subarray of $\bfC$ for given layer and path for writing.
In line~\ref{l:getpointerc-copyc} the array is copied, since the new values in this array may depend on the old value when executing  lines~\ref{l:updatecsc-partsum0}--\ref{l:updatecsc-partsum1}.
We also have functions \call{CGetPtr\_C} and \call{CGetPtr\_T}, which return array pointers for reading, so they do not check \var{RefCount}.

\pairalg{
\caption{\call{CalcT}$(p, \vp)$}
\label{alg:calct}

$t\gets 0$\\
\If{$\vp=0$}{
    \lFor{$\lambda \gets  3 ... m$}{$\call{ProcessPhase}(p,\lambda,0)$} \label{l:calct-proc0}

}
\Else{
    $\lambda\gets m$ \label{l:calct-deepest0}\\
    \While{$\lambda\!\geq\!2 \wedge \vp \text{ is odd} \wedge 1\!<\!\vp< 2^\lambda-1$}{
        $\vp\gets \floor{\frac{\vp-1}{2}}, \lambda\mgets 1$
    } \label{l:calct-deepest1}
    \While{$\lambda \leq m$}{ \label{l:calct-main0}
        $\call{ProcessPhase}(p,\lambda,\vp)$ \label{l:calct-procvp}\\
        $\vp\gets 2\vp+1,\lambda\pgets 1$
    } \label{l:calct-main1}
}
}{
\caption{\call{ProcessPhase}$(p, \lambda ,\psi)$}
\label{alg:clusterops}
$\Omega\gets$ operators for given $\lambda,\psi$ from Tables~\ref{t:op2}--\ref{t:opm}\\
$N\gets 2^{m-\lambda}$, $C\gets\call{CGetPtr\_C}(p,\lambda)$\\
\For{$\omega \in \Omega$}{
    \If{$\omega=\sigma_{i,t,j}$ or $\omega=\osigma_{i,t,j}$}{
        $t'\gets (i+t+j)/2$ \label{l:clusterops-sigma0}\\
        $T\gets\call{GetPtr\_T}(p,\lambda,t)$\\
        $T'\gets \call{CGetPtr\_T}(p,\lambda-1,t')$\\
        \For{$I \in[N]$}{
            $T[I]\!\gets\!\omega(T'[I], T'[I\!+\!N], C[2I..2I\!+\!1])$
        } \label{l:clusterops-sigma1}
    } \ElseIf{$\omega=\mu_{i,t,j}$ or $\omega=\omu_{i,t,j}$}{
        $t'\gets i+t+j$ \label{l:clusterops-mu0}\\
        $T\gets\call{GetPtr\_T}(p,\lambda,t)$\\
        $T'\gets\call{CGetPtr\_T}(p,\lambda,t')$\\
        \For{$I \in [N]$}{%
            $T[I]\gets\omega(T'[I],  C[2I..2I+1])$
        }   \label{l:clusterops-mu1}
    } \ElseIf{$\omega=\Delta\mu_{i,1,j}$}{
        $T\gets\call{GetPtr\_T}(p,m,2)$\\
        $\bfR[p]\gets\omega(T[0],C[0..1])$
    }
}
}


\textbf{Alg.~\ref{alg:calct}}, function \call{CalcT} computes  layers to be updated and local phases on these layers, and performs cluster operations via calling function $\call{ProcessPhase}$
in lines~\ref{l:calct-proc0}, \ref{l:calct-procvp}.

\pairalgr{
\caption{\call{ContinueFrozen}$(\vp)$}
\label{alg:processfrozensymbol}

\For{$p\in\bfA$}{
    $\var{Score}[p]\overset{+}{\gets}\min\set{0,\bfR[p]}$ \label{l:processfrozensymbol-pen}\\
    $C\gets\call{GetPtr\_C}(p,m)$; \label{l:processfrozensymbol-gpc} $C[\vp \bmod 2]\gets 0$
}
}{

\caption{\call{ContinueInfo}$(\vp)$}
\label{alg:processinfosymbol}

$V\gets $ empty array of triples \\
\For{$p\in\bfA$}{ \label{l:processinfosymbol-score0}
    $P[p]\gets \var{Score}[p]-|\bfR[p]|$\\
    $X \gets \mathbf{1}[\bfR[p]<0]$\\
    $(\var{Score}[p], p, X),(P[p], p, X\oplus1) \to V$\label{l:processinfosymbol-score1}\\
}
$l'=\min\set{2\cdot |\bfA|, l}$\\
sort $V$ in descending order by the first component \label{l:processinfosymbol-sort0}\\ 
$V\gets V[0..l'-1]$; $K[0..l-1]\gets 0$ \label{l:processinfosymbol-k0}\\
\lFor{$i\in[l']$}{$K[V[i][1]]\overset{+}{\gets}1$ \label{l:processinfosymbol-k1}}
\lFor{$p\in\bfA$}{\lIf*{$K[p]=0$}{$\call{KillPath}(p)$}} \label{l:processinfosymbol-kill}
$\mA\gets\bfA$\\ 
\For{$p\in\mA$}{  \label{l:processinfosymbol-cont0}
    $X\gets \mathbf{1}[\bfR[p]<0]$\\
    $\call{GetPtr\_C}(p,m)[\vp \bmod 2]\gets X$ \label{l:processfrozensymbol-gpccall0}\\
    \If{$K[p]=2$}{ \label{l:processinfosymbol-clone}
        $p'\gets \call{ClonePath}(p)$; $\var{Score}[p']\gets P[p]$ \label{l:processfrozensymbol-clonecall}\\
        $\call{GetPtr\_C}(p',m)[\vp \bmod 2]\!\gets\! X\!\oplus\! 1$ \label{l:processfrozensymbol-gpccall1}\\        
    }    
} \label{l:processinfosymbol-cont1}

}


\textbf{Alg.~\ref{alg:clusterops}}, function \call{ProcessPhase} performs cluster operators, given for each layer and phase in Tables~\ref{t:op2}--\ref{t:opm}, for path
$p$.
If operator $\omega$ is of type $\sigma$ or $\osigma$, in lines~\ref{l:clusterops-sigma0}--\ref{l:clusterops-sigma1} the operator is executed for clusters on layer $\lambda-1$,
and the results are stored in clusters on layer $\lambda$.
If operator $\omega$ is of type $\mu$, $\omu$ or $\Delta\mu$, in lines~\ref{l:clusterops-mu0}--\ref{l:clusterops-mu1} the operator is executed for $t'$-clusters on layer $\lambda$,
and the results are stored in $t$-clusters on the same layer $\lambda$.
Note that  $t'\neq t$ and pointers $T$ and $T'$ do not point to the same array of clusters.
If $\lambda=m$, this function assigns the output LLR to $\bfR[p]$.

In \textbf{Alg.~\ref{alg:processfrozensymbol}}, function \call{ContinueFrozen} extends all active paths by $u_\vp=0$, and if the output LLR sign for a given path is not consistent with the frozen value (i.e. $L_\vp<0$), then the score of this path is penalized by the absolute value of the  LLR in line~\ref{l:processfrozensymbol-pen}.

In \textbf{Alg.~\ref{alg:processinfosymbol}}, function \call{ContinueInfo} extends each active path from set $|\bfA|$ by two possible values of symbol $\hat u_{\vp}$ and leave maximum of $l$ paths with the highest score.
In lines~\ref{l:processinfosymbol-score0}--\ref{l:processinfosymbol-score1} the scores of $2|\bfA|$ continuations are computed.
Global array $\bfR$ is used to store output LLRs provided by \call{CalcT} function, for path $p$.
Path $p$ is split into two paths: one is continued according to the sign of the output LLR $R[p]$, and its score is equal to the score of the original path $\var{Score}[p]$.
The other is continued by the other value, and its score is penalized by $|R[p]|$, see \eqref{eq:taudef}.
Array $P$ is used to store penalized scores of each path continuation.
Scores, path indices and path continuations are stored as triples in array $V$.
In lines~\ref{l:processinfosymbol-sort0}--\ref{l:processinfosymbol-kill}, this array is sorted, and excess paths with the lowest score are dropped from $V$.

Then, in lines~\ref{l:processinfosymbol-k0}--\ref{l:processinfosymbol-k1}, the number of saved continuations of each path is stored in array $K$.
All active paths that have no continuations in array $V$, are killed via procedure \call{KillPath} in line~\ref{l:processinfosymbol-kill}.
All other active paths are continued by the loop in lines~\ref{l:processinfosymbol-cont0}--\ref{l:processinfosymbol-cont1}.
If a path has only one continuation saved in array $V$, it is continued by the hard decision $X$ on the output LLR, because the other continuation has lower score.
If a path has both of possible continuations in array $V$, it is cloned and the clone is  continued by $X\oplus 1$ in lines~\ref{l:processinfosymbol-clone}--\ref{l:processinfosymbol-cont1}.

\pairalg{
\caption{\call{KillPath}$(p)$}
\label{alg:killpath}

\For{$(\lambda, d)\in[m+1]\times[5]$}{
    $i\gets\var{ArrayIndex}[\lambda][p][d]$\\
    $\var{RefCount}[\lambda][i][d]\overset{-}{\gets}1$\\
    \If{$\var{RefCount}[\lambda][i][d]=0$}{ \label{l:killpath-ref0}
        $\call{Push}(i, \var{FreeArrayIndices}[\lambda][d])$}}\label{l:killpath-ref1}
$\bfA\gets\bfA\setminus\set{p'}$ \label{l:killpath-inact}
}{
\caption{\call{ClonePath}$(p)$}
\label{alg:clonepath}

choose $p'\in [l]\setminus\bfA $\\
\For{$\lambda\in[m+1]$}{$\var{ArrayIndex}[\lambda][p'][0..4]\gets\var{ArrayIndex}[\lambda][p][0..4]$}
$\bfA\gets\bfA\cup\set{p'}$\\
\Return{$p'$}
}

In \textbf{Alg.~\ref{alg:killpath}, \ref{alg:clonepath}} functions \call{KillPath} and \call{ClonePath} are presented.
They are the same as in the straightforward implementation, the only difference is that we deallocate arrays for all cluster dimensions.

Denote the first information and the last frozen symbol indices by $\Phi_0=\min\mI$, $\Phi_1=\max \mF$.
Observe that $L_{\Phi_0}(\bnull^{\Phi_0})$ is a common additive term for all path scores.
Thus, computing LLRs $\tL_\vp$ for $\vp\in[\Phi_0]$ can be skipped.
So, the main loop in lines~\ref{l:decode-main0}--\ref{l:decode-main1} can be started from phase $\Phi_0$. In this case, one needs to precompute some clusters, if they are
used in the first information phase $\Phi_0$.

Observe also that path score can only decrease with the increase of $\vp$.
However, if a phase corresponds to information symbol, the best path is not penalized.
So, the score of the best path during phases $\vp=\Phi_1+1,...,n-1$ is not changed, but the scores of other paths may decrease.
One can obtain the best path $\hat u_0^{\Phi_1}$ on phase $\vp=\Phi_1$ and then switch to SC decoding, not changing the behaviour of the list decoder.
The modifications needed to be done to perform this trick are obvious, so we omit them for the sake of brevity.

\section{Numerical Results}


\pairfig{
\includegraphics[width=\figwidth\textwidth]{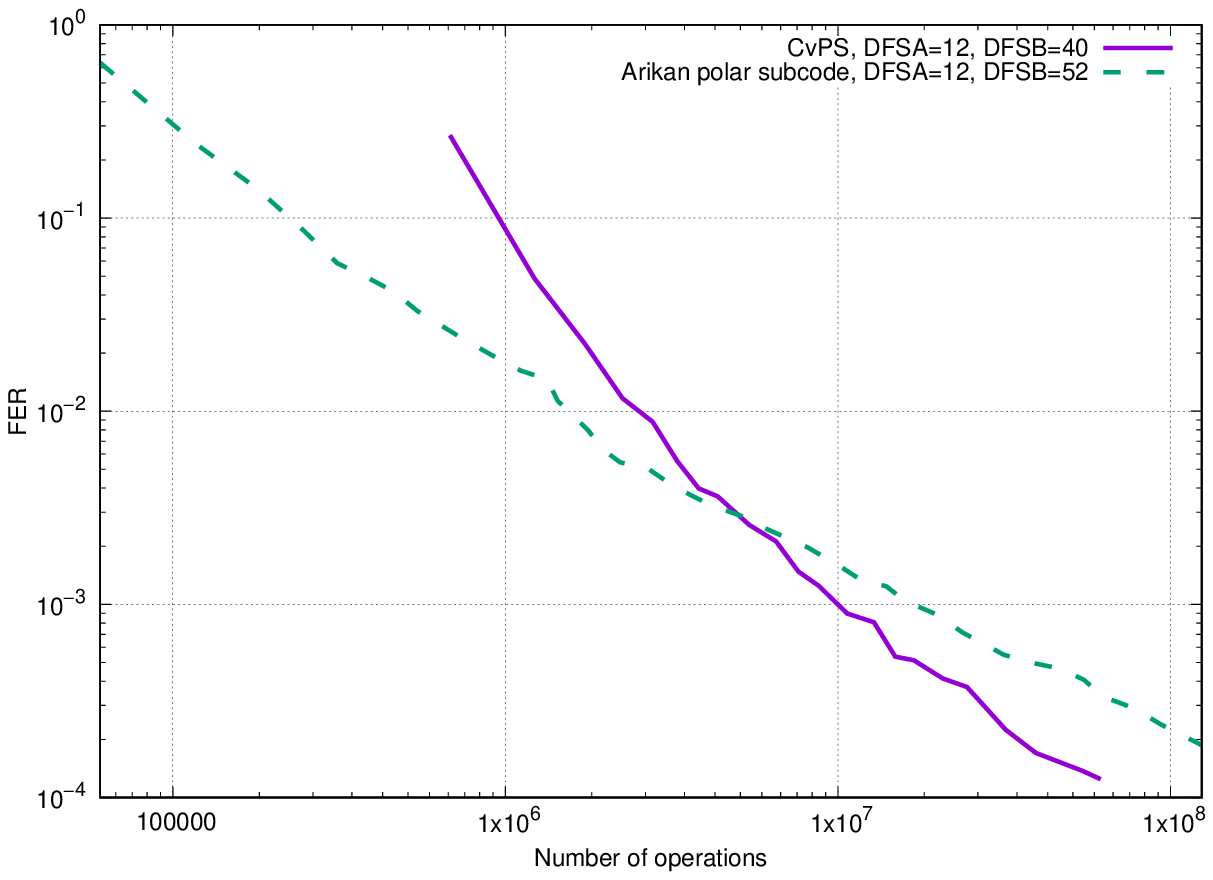}
\caption{$(4096,2048)$ CvPS and Arikan polar subcode.}
\label{fig:cvpcferop40962048}
}{
\includegraphics[width=\figwidth\textwidth]{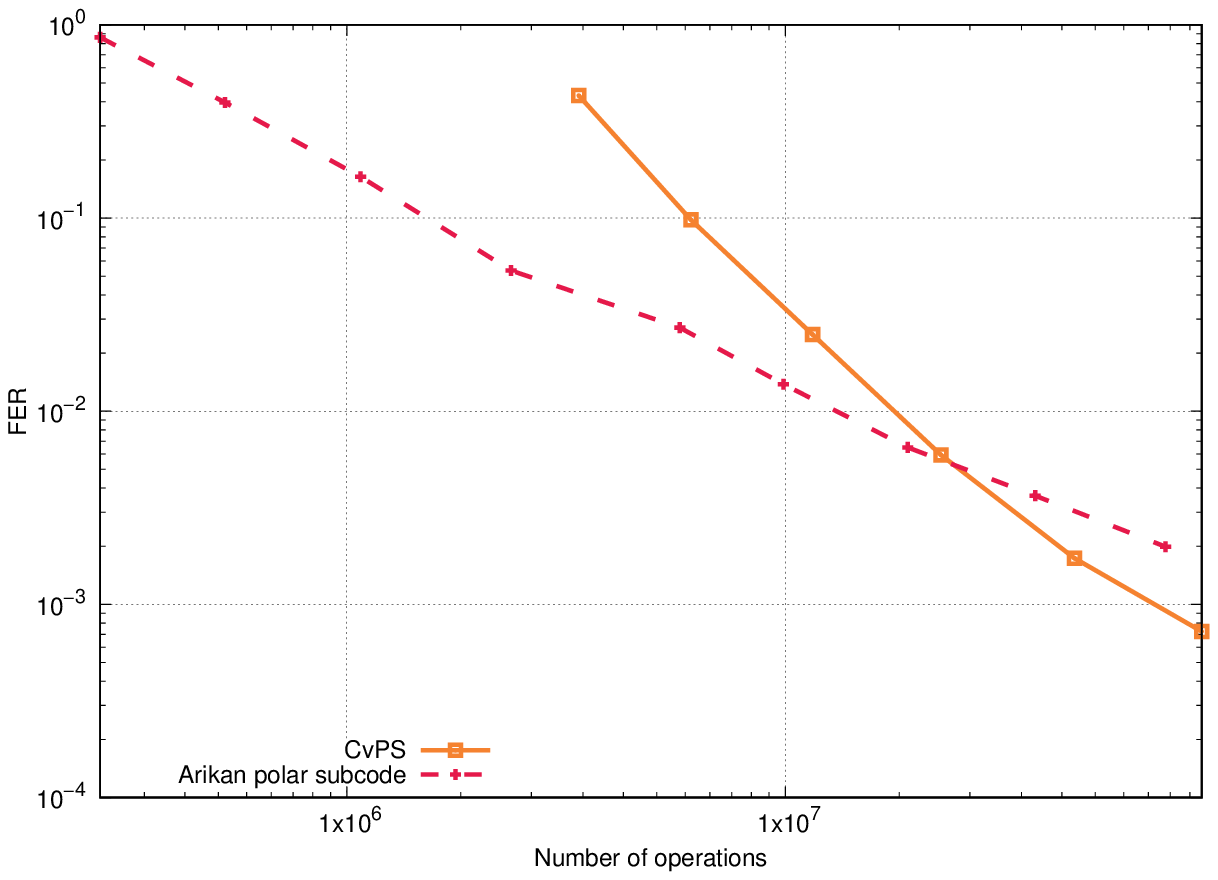}
\caption{$(16384,8192)$ CvPS and Arikan polar subcode.}
\label{fig:cvpcferop16k}
}



In Fig.~\ref{fig:cvpcferop40962048} the FER / complexity curves for a $(4096,2048)$ convolutional polar subcode (CvPS) and Arikan polar subcode are presented.
The CvPS is constructed as shown in \cite{morozov2019distance}, \cite{morozov2019simplified}. 
The simulations were over AWGN channel with BPSK modulation, with SNR\ $E_b/N_0=1.25$ dB.
The intersection point corresponds to $l=96$ for the Arikan polar subcode and $l=11$ for the CvPS.

In Fig.~\ref{fig:cvpcferop16k} the FER / complexity curves for the $(16384,8192)$ CvPS and the Arikan polar subcode are presented.
The SNR of the AWGN channel is $E_b/N_0=1$ dB.
The intersection point corresponds to $l=90$ for the Arikan polar subcode and $l=8$ for the CvPS.

In \cite{saber2018convolutional}, two SC decoding algorithms for CvPCs are proposed: the probability-based and LLR-based.
The probability-based algorithm from \cite{saber2018convolutional} has the same complexity the straightforward SC decoding, presented in Section~\ref{ss:scsf}.
The LLR-based algorithm from \cite{saber2018convolutional} has greater complexity than the presented straightforward SC decoding. 



\section{Summary}
Efficient implementations of SC and list decoding of convolutional polar codes are proposed.
The SC decoding implementation can be used for convolutional polar kernels processing.

The proposed implementation reduces complexity by more than $2$ times compared to the straightforward decoding
of CvPCs.
The complexity reduction is achieved by reusing already computed clusters, by improved implementation of cluster operators, and
by efficient computing of maxima of pairwise sums needed to obtain output clusters.
Further complexity reduction is obtained by skipping the initial block of frozen phases and switching to SC decoding after
processing the last frozen symbol.

\bibliographystyle{IEEETran}

\end{document}